\documentclass[aps,amsmath,amssymb,reprint,superscriptaddress]{revtex4-2}
\usepackage[T1]{fontenc}
\usepackage[latin9]{inputenc}
\usepackage{bm}
\usepackage{amsmath}
\usepackage{amssymb}
\usepackage{babel}
\usepackage{placeins}
\usepackage{xcolor}
\usepackage{graphicx}
\usepackage{hyperref}

\hypersetup{
  colorlinks   = true, 
  urlcolor     = magenta, 
  linkcolor    = magenta, 
  citecolor   = magenta 
}

\begin{document}

\preprint{APS/123-QED}

\title{Magnon-microwave backaction noise evasion in cavity magnomechanics}

\author{V.A.S.V. Bittencourt}%
\email{sant@unistra.fr}
\affiliation{ISIS (UMR 7006), Universit\'{e} de Strasbourg, 67000 Strasbourg, France}

\author{C.A. Potts}
\affiliation{Kavli Institute of NanoScience, Delft University of Technology, PO Box 5046, 2600 GA Delft, Netherlands}
\affiliation{Niels Bohr Institute, University of Copenhagen, Blegdamsvej 17, 2100 Copenhagen, Denmark}
\affiliation{Center for Quantum Devices, Niels Bohr Institute, University of Copenhagen, 2100-DK Copenhagen, Denmark}

\author{J.P. Davis}
\affiliation{Department of Physics, University of Alberta, Edmonton, Alberta T6G 2E9, Canada}

\author{A. Metelmann}
\affiliation{ISIS (UMR 7006), Universit\'{e} de Strasbourg, 67000 Strasbourg, France}
\affiliation{Institute for Theory of Condensed Matter and Institute for Quantum Materials and Technology, Karlsruhe Institute of Technology, 76131, Karlsruhe, Germany}

\begin{abstract}
In cavity magnomechanical systems, magnetic excitations couple simultaneously with mechanical vibrations and microwaves, incorporating the tunability of magnetism and the long lifetimes of mechanical modes. Applications of such systems, such as thermometry and sensing, require precise measurement of the mechanical degree-of-freedom. In this paper, we propose a scheme for realizing backaction evading measurements of the mechanical vibrations in cavity magnomechanics. Our proposal involves driving the microwave cavity with two tones separated by twice the phonon frequency and with amplitudes satisfying a balance relation. We show that the minimum added imprecision noise is obtained for drives centered around the lower frequency magnon-microwave polaritons, which can beat the standard quantum limit at modest drive amplitudes. Our scheme is a simple and flexible way of engineering backaction evasion measurements that can be further generalized to other multimode systems. 
\end{abstract}

\maketitle

\section*{Introduction}

One of the most iconic features of quantum mechanics is the disturbance of a system due to measurements \cite{braginsky1980quantum}. The random nature of quantum measurements  
implies the addition of measurement noise to the system to be probed \cite{clerk2010introduction}. 
We can understand such a backaction effect by considering a simple quantum harmonic oscillator.  
Any position measurement will introduce a disturbance in the momentum, which  
is fed back to the position, manifesting itself as additional noise in future position measurements. Such measurement backaction effects can hinder applications that require precise measurement of an observable, in particular sensing.

Backaction evasion (BAE) schemes provide a pathway
to evade any measurement backaction on an observable. 
This means that the measurement of such a quantum non-demolition (QND) observable, does not disturb its evolution.
The price to be paid is strong noise contamination in non-commuting observables. There are two requirements for a BAE scheme: the QND observable has to commute with itself at different times, and the QND observable has to commute with the interaction Hamiltonian describing the coupling between the measured system and the meter \cite{braginsky1980quantum}. For simple quantum harmonic oscillator, this can be accomplished, for instance, by measurements of the position quadrature as the QND observable, by means of a linear system-meter interaction Hamiltonian. System-meter interactions that yield BAE measurement can be engineered in hybrid systems architecture, a prominent example being the BAE scheme for measuring mechanical vibration via light in optomechanical systems \cite{clerk2008back, woolley2013twomodebackaction, aspelmeyer2014cavity, yanay2017shelvingstyle, hauer2018phononquantumnondemolition, shomroni2019optical}. Other examples include electromechanical systems \cite{hertzberg2010back, lecocq2015quantum, ockeloen2016quantum, liu2022quantumbackaction}, atomic ensembles coupled to mechanical resonators \cite{moller2017quantumbae}, Bose-Einstein condensates \cite{altuntacs2023quantumbaeboseeinstein}, and superconducting qubits coupled to microwave cavities \cite{gambetta2006qubitphoton}.

\begin{figure}[b]
\includegraphics[width = \columnwidth]{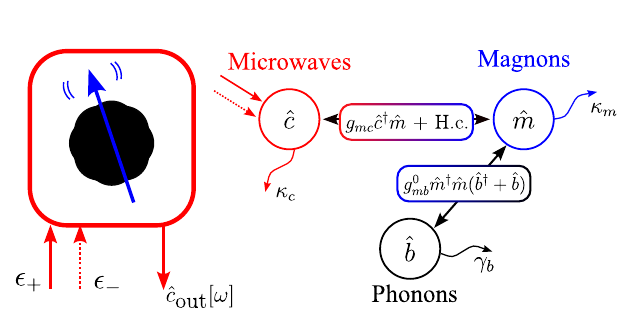}
 \caption{\textbf{Schematic depiction of a cavity magnomechanical system.} (a) The magnetic excitations (blue) couple simultaneously to a microwave mode (red) and mechanical vibrations (black). The cavity can be driven, and the output is used to probe the mechanics. The system is described by a model of interacting bosonic modes subject to dissipation. 
 } 
\label{Fig:Schematic01}
\end{figure}

Cavity magnomechanical systems have recently emerged as a promising platform for quantum technologies \cite{zhang2016cavity,rameshti2022cavity}. In such systems, a magnetic element, usually made of yttrium iron garnet, is loaded into a microwave cavity, as we depict in Fig.~\ref{Fig:Schematic01}. The magnetic excitation (magnons) couple simultaneously to the microwaves (via magnetic dipole coupling) and to the elastic vibrations of the material (via magnetoelastic effects) \cite{lecraw1961extremely,spencer1958magnetoacoustic,zhang2016cavity,Gonzalez_2020_Theory}. Typically, magnons and microwaves couple strongly and form hybrid magnon-microwave polaritons with frequencies defined by the coupling strength and the tunable magnon frequency. Such a system allows the drive and measurement of phonons via the microwave resonator while retaining the tunability of the magnons, as well as the use of the intrinsic magnon-microwave hybridization, for example, for efficient cooling and amplification of the mechanics \cite{potts2021dynamical}. Among the potential applications proposed for such systems are the generation of entangled states \cite{li2018magnon, li2021entangling, fan2022microwaveopticsentanglement}, the generation of squeezing of magnons and phonons via magnon nonlinearities \cite{li2021squeezing, ding2022magnonsqueezing}, and noise-based thermometry \cite{potts2020magnon}. The aforementioned applications require precise measurement of the mechanics and thus can be hindered by measurement backaction, an issue that has not been addressed in the field.

In this paper, we propose a BAE scheme for measuring phonon quadratures in a cavity magnomechanical system. Our scheme exploits the radiation pressure-like coupling between the magnons and the mechanical mode to realize
a QND Hamiltonian for a mechanical amplitude quadrature. The QND Hamiltonian is engineered by modulating the force driving the elastic vibrations due to magnetostriction, akin to BAE schemes proposed and implemented in opto and electromechanical systems \cite{clerk2008back,woolley2013twomodebackaction,hertzberg2010back}. Such modulation is accomplished by driving the microwave mode with two coherent tones, with balanced amplitudes. We then show that different hybrid mode separations yield different performances of the BAE scheme in terms of imprecision noise added to the mechanical measurement.  
Crucially, at a reasonable drive amplitude, it is possible to beat the standard quantum limit of added noise. To study the robustness of the system we investigate in depth imperfections of the parameters.  
Our results present a simple yet robust framework for performing BAE measurements of mechanics in a cavity magnomechanical system, but they can be adapted to other multimode systems in which the measured system couples to a composite system exhibiting hybridization, for instance, optomechanical systems with two coupled optical cavities. Furthermore, the BAE scheme can be combined with other proposals, such as the use of nonlinearities for enhancing force measurements \cite{zhang2024quantumweak}.

The paper is structured as follows. We first present a brief introduction to cavity magnomechanics and quantify the backaction noise in a single-tone drive setup, focusing on the framework of the recently demonstrated dynamical backaction evasion \cite{potts2022dynamical}. We then introduce our BAE scheme and compute the imprecision noise added to the measurement of the BAE quadrature via the output of the microwave mode. Finally, we quantify the robustness of the BAE scheme under imperfect conditions, focusing on the changes in the added noise to the measurement of the mechanical quadrature. We outline the derivations and present important formulas in the supplementary information (SI) \cite{SI}.

\section*{Results}

\subsection*{Cavity magnomechanics and backaction noise}
\label{sec:Magnomech}

In this section, we introduce cavity magnomechanical systems and the effects of backaction noise on the measurement of the phonon mode in standard single-tone drive setups, which will provide a framework for the main core of our results.

The cavity magnomechanical system depicted in Fig.~\ref{Fig:Schematic01} can be modelled with the following Hamiltonian \cite{zhang2016cavity, Gonzalez_2020_Theory,bittencourt2023dynamical}
\begin{equation}
\label{eq:simphamil}
\begin{aligned}
\frac{\hat{H}}{\hbar} &= \omega_c \hat{c}^\dagger \hat{c} + \omega_m \hat{m}^\dagger \hat{m} +\omega_b \hat{b}^\dagger \hat{b}+\frac{\hat{H}_{\rm{drive}}}{\hbar} \\
&+ g_{mc} \left( \hat{m}^\dagger \hat{c} + \hat{m} \hat{c}^\dagger \right) +g_{mb}^0 \hat{m}^\dagger \hat{m}\left(\hat{b}^\dagger +\hat{b} \right),
\end{aligned}
\end{equation}
where $\hat{H}_{\rm{drive}}$ denotes a coherent microwave drive term. A microwave mode $\hat{c}$ with frequency $\omega_c$ couples to a magnon mode $\hat{m}$ with frequency $\omega_m$, which in turn couples to a phonon mode $\hat{b}$ with frequency $\omega_b$. The magnon frequency $\omega_m$ can be tuned by an applied external field \cite{stancil2009spinwaves}. Magnons and microwaves couple via magnetic dipole interaction, which yields the coupling rate $g_{mc}$. At the same time, magnetoelastic effects \cite{callen1968magnetostriction} are responsible for coupling magnons and phonons with the single magnon coupling rate $g_{mb}^0$. We consider exclusively the coupling between the uniform magnon mode (the Kittel mode), with a single cavity mode and a single low-frequency phonon mode. A more detailed discussion and derivation of the coupling rates can be found in Refs.~\cite{Gonzalez_2020_Theory, bittencourt2023dynamical}.  
The numerical value of the parameters that we will use throughout the paper are summarized in table \ref{Table0} and correspond to experiments in which a 3D microwave cavity is loaded with a YIG sphere with radius $\sim100\,\mu$m.

The microwave-magnon coupling can be stronger than both microwave and magnon decay rates, $\kappa_c$ and $\kappa_m$ respectively, generating a hybridization between the modes \cite{huebl2013high,zhang2014strongly,potts2020strong}. In such a regime, two polariton modes form at frequencies $\omega_{\pm}$. We call the mode with frequency $\omega_+>\omega_-$ the upper hybrid mode while the other is the lower hybrid mode. In the case where the magnon mode is resonant with the microwave mode $\omega_m = \omega_c$, the difference between the hybrid modes frequencies is $\sim 2 g_{mc}$. Throughout the paper, we will consider exclusively the situation in which magnons and microwaves are at resonance, corresponding to maximum hybridization between the modes.

\begin{figure}[t]
\includegraphics[width = \columnwidth]{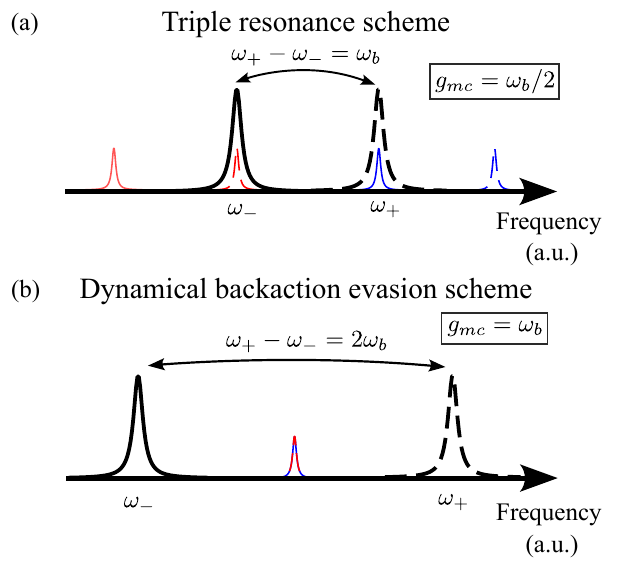}
\caption{\textbf{Different cavity magnomechanics schemes.} The Lorentzians indicate the resonance frequency and linewidths of the hybrid modes as they would be measured, for example, via transmission. Relevant mechanical sidebands are also indicated.  (a) Triple resonance scheme: the magnon-microwave coupling $g_{mc}$ is half of the phonon frequency. The blue (red) sideband of the lower (upper) hybrid mode coincides with the upper (lower) hybrid mode frequency. (b) Dynamical backaction evasion scheme: the magnon-microwave coupling is equal to one phonon frequency. In this scheme, the sidebands of the hybrid modes coincide.} 
\label{Fig:Schematic02}
\end{figure}

The combination of the hybridization with the sidebands generated by the magnomechanical interaction, gives a unique trait of the system in which the sidebands can lie close to the hybrid modes frequencies or overlap. This is depicted in Fig.~\ref{Fig:Schematic02} for the two configurations which we call triple resonance (modes split by one phonon frequency) and the dynamical backaction evasion schemes (modes split by twice phonon frequency), that have been explored in previous works \cite{potts2021dynamical,potts2022dynamical,bittencourt2023dynamical}. These different configurations can be used for enhancing or balancing Stokes and anti-Stokes processes, yielding efficient cooling and amplification of the phonon mode \cite{potts2021dynamical}, or dynamical backaction evasion \cite{potts2022dynamical,bittencourt2023dynamical}.

\begin{table*}[t!]
\caption{Typical parameters of a cavity magnomechanical system consisting of a magnetic sphere loaded in a 3D microwave cavity.
\label{Table0}}
\begin{tabular}{ |c|c|c| } 
\hline
\textbf{Parameter} & \textbf{Symbol} & \textbf{Value} \\
\hline
Microwave mode frequency & $\omega_{c}$ &  $2 \pi \times 10$ GHz \\
\hline
Magnon mode frequency & $\omega_{m}$ & $ \omega_c$ \\
\hline
Phonon mode frequency & $\omega_{b}$ & $10^{-3} \omega_c$ \\
\hline
Microwave total decay rate & $\kappa_c$ & $2 \times 10^{-4} \omega_c$ \\
\hline
Magnon mode decay rate & $\kappa_{m}$ & $10^{-4} \omega_c$ \\
\hline
Phonon intrinsic decay rate & $\gamma_b$ & $10^{-7} \omega_c$ \\
\hline
Magnon-microwave coupling rate & $g_{mc}$ & $\sim 10^{-3} \omega_c$ (different cases will be considered) \\
\hline
Magnomechanical vacuum coupling rate & $g_{mb}^{(0)}$ & $10^{-12} \omega_c$\\
\hline
Microwave drive amplitude & $\epsilon_d$ & in the range $\{10^3, 10^6\}\times \sqrt{\omega_c}$ \\
\hline
\end{tabular}
\end{table*}

The measurement of the phonon mode can be accomplished by monitoring the output field of the microwave mode. Since the noise spectrum of the output field has imprints of the mechanical spectrum, the backaction noise driving the mechanics will unavoidably contaminate the measurement \cite{aspelmeyer2014cavity,clerk2010introduction}. We describe this effect via the standard input-output relation for the microwave mode $\hat{c}_{\rm{out}}[\omega] = \hat{c}_{\rm{in}}[\omega] - \sqrt{\kappa_c} \hat{c}[\omega].$ The operator $\hat{c}[\omega]$ describe the fluctuations of the microwave cavity mode on top of a coherent state generated by the strong drive. It can be obtained by solving the linearized Heisenberg-Langevin equations in the frequency domain, which is shown in detail in \cite{bittencourt2023dynamical} and in the SI \cite{SI}. We can compute the frequency-symmetrized noise spectrum of a general quadrature
\begin{equation}
\hat{x}_{\theta, \rm{out}}[\omega] = \frac{\hat{c}_{\rm{out}}[\omega] e^{-i \theta} +\hat{c}_{\rm{out}}^\dagger[\omega] e^{i \theta}}{\sqrt{2}},
\end{equation}
given by $\bar{S}_{\theta \theta}[\omega] = (S_{\theta \theta}[\omega] +S_{\theta \theta}[-\omega])/2$, where
\begin{equation}
\label{eq:NoiseSpectrumOneTone}
S_{\theta \theta}[\omega] = \frac{1}{2}\int_{-\infty}^\infty \frac{d \omega}{ 2 \pi} \langle \hat{x}_{\theta, \rm{out}}[\omega],  \hat{x}_{\theta, \rm{out}}[\omega^\prime] \rangle. 
\end{equation}
Typically, $\bar{S}_{\theta \theta}[\omega]$ is measured via a homodyne setup, where $\theta$ is tuned by a local field which is mixed with the output signal. The output noise spectrum yields
\begin{equation}
\label{eq:noisemeas}
\begin{aligned}
\bar{S}_{\theta \theta}[\omega] &= \vert \mathcal{G}_{\theta} [\omega] \vert^2 \left[\bar{S}_{xx}^{(0)}[\omega] + \bar{S}_{\rm{BA}}[\omega] + \bar{S}_{\rm{imp}} [\omega] \right]. 
\end{aligned}
\end{equation}
The goal of such a measurement is to probe $\bar{S}_{xx}^{(0)}[\omega]$, the uncoupled noise spectrum of the phonon mode, or mechanical noise spectrum, which is given by
\begin{equation}
 \bar{S}_{xx}^{(0)}[\omega] =  \frac{\gamma_b}{2} \left(2 n_{b} + 1 \right) \left( \vert \chi_b[\omega] \vert^2+ \vert \chi_b[-\omega] \vert^2 \right),  
\end{equation}
where $n_b$ is the phonon bath occupancy, and $\chi_b[\omega] = 1/(-i(\omega - \omega_b) + \gamma_b/2)$ is the phonon mode susceptibility. $\bar{S}_{xx}^{(0)}[\omega]$ has units of inverse frequency and gives the position noise spectrum of the mechanical vibration via $x_{\rm{ZPF}}^2  \bar{S}_{xx}^{(0)}[\omega]$, where $x_{\rm{ZPF}}$ are the zero-point fluctuations of the phonon mode. For a magnomechanical system, $x_{\rm{ZPF}}$ depends on the elastic coefficients of the material as well as the particular mode profile of the elastic vibrations \cite{bittencourt2023dynamical}. 
The gain $\vert \mathcal{G}_{\theta} [\omega] \vert^2$ depends on the quadrature being measured. For the setup considered here, the maximum gain is obtained for $\theta = \pi/2$. The backaction noise $\bar{S}_{\rm{BA}}[\omega]$ is a consequence of the non-QND character of the quadrature being measured. We provide explicit expressions for all the terms appearing in Eq.~\eqref{eq:noisemeas} in the SI \cite{SI}.

In addition to its thermal motion,
the phonon mode experiences a displacement in response to the backaction noise. Such a displacement contaminates the measurement, manifesting as additional (backaction) noise added on top of the imprecision noise. In a typical optomechanical system with a single optical mode, the minimum added noise at the mechanical frequency, referred to as the standard quantum limit (SQL), is
\begin{equation}
\bar{S}_{\rm{BA}}^{\rm{SQL}}[\omega_b] + \bar{S}_{\rm{imp}}^{\rm{SQL}} [\omega_b] = \bar{S}_{xx}^{(0)}[\omega_b],
\end{equation}
with the backaction and imprecision noise having equal contributions. Such minimum can be achieved by tuning the drive amplitude \cite{aspelmeyer2014cavity}, which we indicate by $\epsilon_d$. The SQL can also be written in terms of added quanta defined by
\begin{equation}
n_{\rm{BA}} = \frac{\gamma_b}{4}\bar{S}_{\rm{BA}}[\omega_b], \quad n_{\rm{imp}} = \frac{\gamma_b}{4}\bar{S}_{\rm{imp}}[\omega_b],
\end{equation}
such that the total added quanta to the measurement is
\begin{equation}
n_{\rm{add}} = n_{\rm{BA}} + n_{\rm{imp}}.
\end{equation}
For a drive power achieving the SQL, we have
\begin{equation}
\label{eq:totaddedSQL}
n_{\rm{add}}^{\rm{SQL}} = n_{\rm{BA}}^{\rm{SQL}}+n_{\rm{imp}}^{\rm{SQL}} = \frac{1}{2}.
\end{equation}
In other words, in the best case, half a quanta is added to the measurement of the phonon mode. 

\begin{figure}[h]
\includegraphics[width =  1.0 \columnwidth]{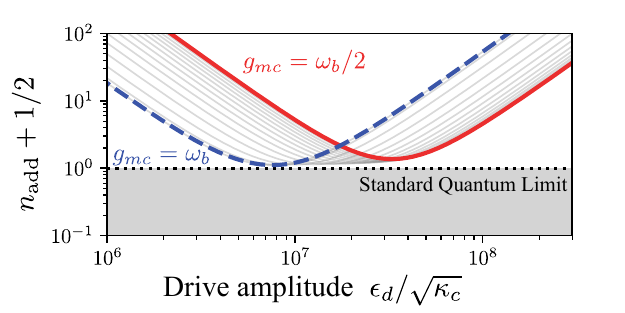}
\caption{\textbf{Number of added quanta to the measurement.} The number of added quanta is offset by the vacuum fluctuations in a single-tone cavity magnomechanical system at zero temperature and for a zero detuned drive as a function of the drive amplitude $\epsilon_d/\sqrt{\kappa_c}$. The curves shown are for magnon-microwave couplings varying from $\omega_b/2$ (red, continuous curve) to $\omega_b$ (blue dashed line), with the thin lines corresponding to intermediate values. The values of drive amplitude depicted here correspond to linear magnomechanical couplings $\vert g_{mb} \vert$ in the range $\approx 10^{-3} \kappa_c$ to $0.5 \kappa_c$. All other parameters as in Table \ref{Table0}.} 
\label{Fig:AddednoiseStd}
\end{figure}

The hybridization between magnons and microwaves can prevent such a measurement from reaching the standard quantum limit, as the interplay between imprecision and backaction noise depends on the magnon-microwave coupling. This is shown in Fig.~\ref{Fig:AddednoiseStd}, which depicts the total added quanta to the measurement $n_{\rm{add}}$ as a function of the drive amplitude $\epsilon_d$. The drive frequency $\omega_d$ is fixed at the magnon/microwave frequency $\omega_{c} = \omega_m$, since this choice eliminates dynamical backaction effects \cite{potts2022dynamical,bittencourt2023dynamical}. The imprecision noise decreases with the drive amplitude, while the backaction noise increases with the drive amplitude. Their interplay yields the minimum in Eq.~\eqref{eq:totaddedSQL} indicated by the dotted line in the figure. Since the measurement of the mechanics is performed via the microwaves instead of the directly coupled magnon mode, weaker magnon-microwave couplings yield smaller gains and, thus, larger imprecision noise, but also less backaction noise due to the smaller (linear) magnomechanical coupling. As a result, the minimum of the added noise for weaker magnon-microwave couplings is achieved at a higher drive amplitude than that for stronger magnon-microwave couplings. 

\subsection*{Scheme for backaction noise evasion}
\label{sec:qndmagnomech}

In the last section, we have pointed out that the backaction noise driving the phonon mode is due to the non-QND nature of the position quadrature $\hat{x}_b$. To implement measurements that avoid backaction noise, we consider two requirements \cite{braginsky1980quantum}: the observable to be measured has to commute with itself at all times, thus preventing the dynamics of the system from contaminating future measurements, and the measured observable has to commute with the system-meter interaction term, which ensures that no noise from the meter is fed into the system and then fed back to the measurement. This corresponds to a BAE measurement, which is also called a QND measurement.

\begin{figure}[t]
\includegraphics[width =  \columnwidth]{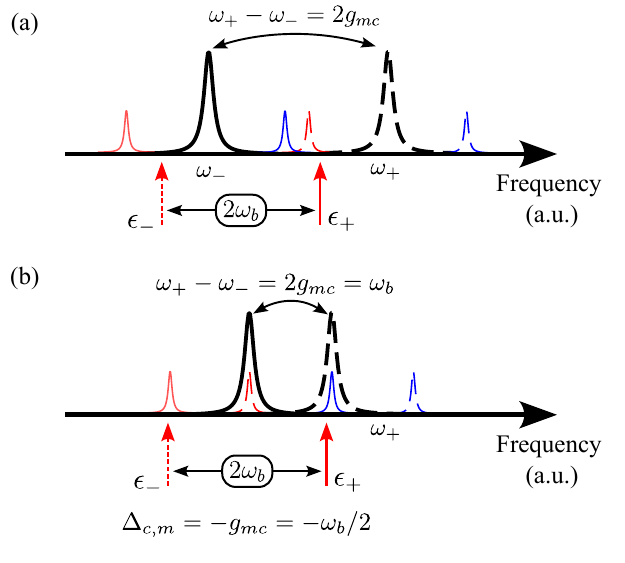}
\caption{\textbf{Frequency configuration for the backaction evasion scheme.} (a) General framework: the two tones have to be separated by twice the phonon frequency, and the drive amplitudes $\epsilon_{\pm}$ have to satisfy Eq.~\eqref{eq:DriveAmplitudes}. The hybrid mode splitting can be arbitrary; (b) one particular configuration in which the mode splitting is set to the triple resonance scheme. In this case, the tones are centered around the lower hybrid mode. The high-frequency tone drives simultaneously the blue sideband of the lower hybrid mode and the upper hybrid mode. Such a configuration yields the minimum imprecision noise in the measurement of the BAE quadrature via the microwave output.} 
\label{Fig:scheme03}
\end{figure}

To engineer a QND Hamiltonian for a quadrature of the phonon mode in a cavity magnomechanical system, we consider a two-tone microwave drive
\begin{equation}
\label{eq:2tonedrive}
\frac{\hat{H}_{\rm{drive}}}{\hbar} = i \sqrt{\kappa_e} (\epsilon_{-} e^{- i \delta t} + \epsilon_{+} e^{i \delta t})e^{i \omega_d t} \hat{c} + \rm{H.c.}, 
\end{equation}
in which the coherent tones $\pm$ have frequencies $\omega_d \pm \delta$. To obtain a QND Hamiltonian, the drives should induce a modulation of the effective force $\propto \hat{m}^\dagger \hat{m}$ (see Eq.~\eqref{eq:simphamil}) driving the phonon mode. This can be accomplished under two conditions: the frequency tone separation $2 \delta$ has to match twice the mechanical frequency $2 \omega_b$, as shown in Fig.~\ref{Fig:scheme03}, and the drive amplitudes have to balance to guarantee that both drives yield the same (linear) magnomechanical coupling. The last requirement implies the following relation
obtained which is obtained from the solution of the classical equations
\begin{equation}
\label{eq:DriveAmplitudes}
\frac{\vert \epsilon_{+} \vert}{\vert \epsilon_{- } \vert}= \frac{\vert \left( i (\Delta_m + \omega_b) - \frac{\kappa_m}{2} \right) \left( i (\Delta_c + \omega_b) - \frac{\kappa_c}{2} \right)  + g_{mc}^2 \vert }{\vert \left( i (\Delta_m - \omega_b) - \frac{\kappa_m}{2} \right) \left( i (\Delta_c - \omega_b) - \frac{\kappa_c}{2} \right)  + g_{mc}^2 \vert},
\end{equation}
in which $\Delta_{c,m} = \omega_d - \omega_{c,m}$ is the detuning between the central frequency of the drives $\omega_d$ and the magnon/microwave frequencies. More detailed calculations are provided in the SI \cite{SI}. Such a strategy to engineer a BAE measurement of the mechanics is akin to the one proposed and implemented in opto and electromechanical system \cite{clerk2008back,woolley2013twomodebackaction,hertzberg2010back,lecocq2015quantum}.

\begin{figure}[h]
\includegraphics[width =  \columnwidth]{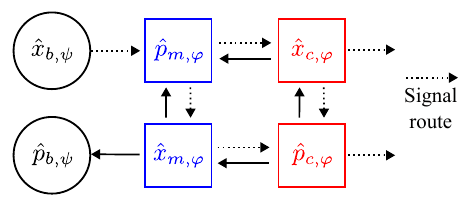}
\caption{\textbf{Flowchart schematizing the phonon signal's route in a magnomechanical system under the two-tones QND scheme.} We have indicated the route that the signal of the QND mechanical quadrature $\hat{x}_{b,\psi}$ takes with dotted arrows. For a vanishing detuning $\Delta_c = 0$, the coupling between orthogonal magnon-microwave quadratures vanishes.} 
\label{Fig:FC02}
\end{figure}

The two aforementioned conditions, the tones separation by twice the phonon frequency and the drive amplitudes balance in Eq.~\eqref{eq:DriveAmplitudes}, engineer an effective modulation of the magnomechanical force driving the phonon mode. This in turn yield the linearized Hamiltonian of the system in the interacting frame with respect to $\omega_b \hat{b}^\dagger \hat{b}$ is
\begin{equation}
\label{eq:qndqadrat}
\begin{aligned}
\frac{\hat{H}_{\rm{QND}}}{\hbar}  &= -\frac{\Delta_c}{2} (\hat{x}_{c, \varphi}^ 2 + \hat{p}_{c, \varphi}^ 2)  -\frac{\Delta_m}{2}  (\hat{x}_{m, \varphi}^ 2 + \hat{p}_{m, \varphi}^ 2) \\
&+ g_{mc} \left( \hat{x}_{m, \varphi} \hat{x}_{c, \varphi} + \hat{p}_{c, \varphi}\hat{p}_{m, \varphi}\right) + 2 G \hat{x}_{m,\varphi} \hat{x}_{b,\psi}.
\end{aligned}
\end{equation}
where the quadratures are defined as ($\alpha = c,m,b$)
\begin{equation}
\begin{aligned}
\hat{x}_{\alpha,\phi} &= \frac{e^ {i \phi} \hat{\alpha}^\dagger + e^ {-i \phi} \hat{\alpha}^\dagger}{\sqrt{2}}, \\
\hat{p}_{\alpha,\phi} &= i\frac{e^ {i \phi} \hat{\alpha}^\dagger - e^ {-i \phi} \hat{\alpha}}{\sqrt{2}},
\end{aligned}
\end{equation}
and the phases $\varphi$ and $\psi$ depend on the relative phases of the coherent steady-state of the magnon mode under the two-tone drive, whose amplitude defines the magnomechanical coupling $G$. The Hamiltonian \eqref{eq:qndqadrat} is QND with respect to the phonon quadrature $\hat{x}_{b, \psi}$. By tuning the relative phase of the drive amplitudes, it is possible to change the phase $\psi$ and thus to alter which phonon quadrature is QND. Such a quadrature is completely unaffected by the coupling to the magnons and by any measurement process done in the magnon-microwave part of the system. Nevertheless, due to the hybridization between magnons and microwaves, the quadrature $\hat{x}_{m,\varphi}$, through which one would measure $\hat{x}_{b, \psi}$ is not a QND observable. This can be seen by analyzing the flowchart in Fig.~\ref{Fig:FC02}, which illustrates the information flow in the system, for example, the magnon quadrature $\hat{p}_{m,\varphi}$ measures the phonon quadrature $\hat{x}_{b,\psi}$, passing the information to the quadratures $\hat{x}_{c,\varphi}$ and $\hat{x}_{m,\varphi}$. No quadrature feeds $\hat{x}_{b, \psi}$, since such a phonon quadrature is a QND observable. Otherwise, no magnon/microwave observable is QND.
To note is, that the QND-nature of the mechanical quadrature $\hat{x}_{b,\psi}$ is as well preserved in the polariton basis. 

The solutions of the equations of motion for $\hat{x}_{b, \psi}$ and $\hat{p}_{b, \psi}$ in frequency domain are
\begin{equation}
\label{eq:QNDquadratures}
\begin{aligned}
\hat{x}_{b, \psi}[\omega] &= \sqrt{\gamma_b} \hat{x}_{b,\psi, {\rm{in}}} [\omega], \\
\hat{p}_{b, \psi}[\omega] &= \sqrt{\gamma_b} \hat{p}_{b,\psi, {\rm{in}}} [\omega] - 2 G \chi_b[\omega] \hat{\xi}_{{\rm{BA}}, x_{\varphi}}[\omega] \\
&\quad + 2 \sqrt{\gamma_b}\Sigma_b[\omega] \chi_b[\omega] \hat{x}_{b,\psi, {\rm{in}} } [\omega].
\end{aligned}
\end{equation}
The only term appearing in the equation for $\hat{x}_{b, \psi}[\omega]$ is the intrinsic (thermal and vacuum) noise driving the mechanics. Otherwise, all the backaction noise is dumped into the orthogonal quadrature $\hat{p}_{b, \psi}[\omega]$, which causes  amplification of the noise of that quadrature. The noise terms $\hat{x}_{b,\psi, {\rm{in}}} [\omega]$ and $\hat{p}_{b,\psi, {\rm{in}}} [\omega]$ describe the intrinsic vacuum and thermal noise acting in the phonon mode, they have the correlations
\begin{equation}
\begin{aligned}
\langle \hat{x}_{b,\psi, {\rm{in}}} [\omega] \hat{x}_{b,\psi, {\rm{in}}} [\omega^\prime] \rangle &= \langle \hat{p}_{b,\psi, {\rm{in}}} [\omega] \hat{p}_{b,\psi, {\rm{in}}} [\omega^\prime] \rangle \\&=  \pi \chi_b^2[\omega]\left(2 n_b + 1 \right) \delta[\omega + \omega^\prime],\\
\langle \hat{x}_{b,\psi, {\rm{in}}} [\omega] \hat{p}_{b,\psi, {\rm{in}}} [\omega^\prime] \rangle &=-\langle \hat{p}_{b,\psi, {\rm{in}}} [\omega] \hat{x}_{b,\psi, {\rm{in}}} [\omega^\prime] \rangle\\
&= i \pi \chi_b^2[\omega]  \delta[\omega + \omega^\prime],
\end{aligned}
\end{equation}
where $\delta$ is the Dirac delta. In the equation for $\hat{p}_{b, \psi}$, $\Sigma_b[\omega]$ is called the phonon self-energy \cite{potts2020magnon}, given by
\begin{equation}
\label{eq:phononselfenergyfull}
\Sigma_{b} [\omega] = i G^2 \left( \Xi_m[\omega + \omega_b] - \Xi_m^*[-\omega - \omega_b] \right),
\end{equation}
where 
\begin{equation}
\Xi_m^ {-1}[\omega] = \chi_m^{-1}[\omega] +g_{mc}^2 \chi_c [\omega].
\end{equation}
is a modified magnon susceptibility. We have indicated the magnon and microwave susceptibilities by $\chi_{c,m } [\omega] = 1/(-i(\omega - \Delta_{m,c}) + \kappa_{m,c}/2)$.
The backaction noise operator $\hat{\xi}_{{\rm{BA}}, x_{\varphi}}[\omega]$ has correlations
\begin{equation}
\begin{aligned}
\langle \hat{\xi}_{{\rm{BA}}, x_{\varphi}}[\omega] \hat{\xi}_{{\rm{BA}}, x_{\varphi}}[\omega^\prime] \rangle &= \pi \Xi_m[\omega] \Xi_m^*[\omega^\prime] (\kappa_m \\ &+ g_{mc} \kappa_c \chi_c[\omega] \chi^*_c[\omega^\prime])\delta[\omega+\omega^\prime],
\end{aligned}
\end{equation}
where we have assumed that the magnon and microwave modes are at zero temperature. 

Finally, unlike dynamical backaction evasion, the BAE scheme does not require the detunings to vanish. Nevertheless, to obtain Eq.~\eqref{eq:qndqadrat}, we have applied a rotating wave approximation 
(RWA) and discarded terms rotating at twice the phonon frequency.  
For the parameters considered here and at zero detuning, the corrections that the counter-rotating terms introduce in the integral of the phonon noise spectrum are $\sim 10^{-4}$ and can be safely ignored \cite{SI}. At finite detunings, the analysis is more involved, but, since at all powers considered here the linearized magnomechanical coupling $\vert G \vert \ll \omega_b$, we expect that the RWA is a good approximation and thus that the same conclusions hold.

\subsection*{Microwave output spectrum and imprecision noise}
\label{sec:output}

We now turn our attention to the measurement of the QND quadrature via the output microwave signal. For this, we consider the standard input-output relation and compute the noise spectrum of a generalized output quadrature. In the present case, we can show that in a small bandwidth around zero frequency, the output noise spectrum is given by \cite{SI}
\begin{equation}
\begin{aligned}
\bar{S}_{\theta \theta}[\omega] &= \vert \mathcal{G}_{x, \theta}[\omega] \vert^2 \left( \bar{S}^{(0)}_{xx}[\omega] + \frac{4 n_{\rm{imp} , \theta}}{\gamma_b} \right),
\end{aligned}
\end{equation}
where $n_{\rm{imp}, \theta}$ is an effective number of quanta added to the measurement of the mechanical spectrum due to the imprecision noise. Since the scheme is BAE, there is no noise added by backaction, compared with Eq. \eqref{eq:noisemeas}.

The imprecision noise depends on the ratio between the uncoupled output spectrum and the gain factor evaluated at zero frequency
\begin{equation}
n_{\rm{imp}, \theta} \propto \frac{\bar{S}_{\rm{out}}^{(0)}[0]}{\vert \mathcal{G}_{x, \theta}[0] \vert^2},
\end{equation}
where $\bar{S}_{\rm{out}}^{(0)}[0]$ is the output spectrum in the absence of the magnomechanical coupling, which exhibits two minima for drives centered around the hybrid modes frequencies $\omega_\pm = \omega_c \pm 2 g_{mc}$. 
In conjunction with a minimal $\bar{S}_{\rm{out}}^{(0)}[0]$, the impression noise is suppressed at large gain, which can be achieved at optimized detunings.  
However, the gain factor for arbitrary detuning can be written as
\begin{equation}
\begin{aligned}
\vert \mathcal{G}_{x, \theta}[0] \vert^2 &= 2 \kappa_c  \vert G \vert^2 g_{mc}^2 \Bigg( \frac{1}{\vert \lambda_+ \vert^2 \vert \lambda_- \vert^2} 
+ 2 {\rm{Re}} \left[\frac{e^{2 i (\varphi - \theta)}}{\lambda_+^2 \lambda_-^2} \right] \Bigg),
\end{aligned}
\end{equation}
where $\lambda_\pm = i (\omega_d - \omega_{\pm}) - (\kappa_c + \kappa_m)/4$. This gain factor is maximized for drives centered around the hybrid modes $\omega_d = \omega_\pm$, or, correspondingly, a detuning $\Delta_c = \mp g_{mc}$. In these cases
\begin{equation}
n_{\rm{imp}, \theta} = \frac{\gamma_b}{4} \frac{\kappa_{{\rm{eqv}}, \theta}}{32 G^2}.
\end{equation}
In this form, the imprecision noise has the same expression as for a single-cavity optomechanical system \cite{clerk2008back}, with an equivalent decay $\kappa_{{\rm{eqv}}, \theta}$ given by
\begin{equation}
\kappa_{{\rm{eqv}}, \theta} =  \frac{\left(g_{mc}^2 (\kappa_c+\kappa_m)^2 + \frac{\kappa_c^2 \kappa_m^2}{4} \right)^2}{\kappa_c g_{mc}^ 2 \left(\frac{\kappa_c \kappa_m}{4} \cos(\varphi - \theta) \pm g_{mc} \kappa_\pm \sin(\varphi - \theta) \right)^ 2}.
\end{equation}
The output measurement angle $\theta_{\rm{opt}}$ that minimizes the above expression is
\begin{equation}
\tan(\varphi - \theta_{\rm{opt}}) = \pm \frac{4g_{mc} (\kappa_c+\kappa_m)}{\kappa_c \kappa_m}.
\end{equation}
The quadrature that optimizes the measurement gain depends on the magnon-microwave coupling and the decay rates, and it is, therefore, related to the hybridization. In fact, by inspecting the flowchart in Fig. \ref{Fig:FC02}, we notice that the signal of the mechanical quadrature distributes among the microwave quadratures due to the coupling between microwave and magnons. For the optimum measurement quadrature, the minimum added quanta added to the measurement is
\begin{equation}
n_{\rm{imp,opt}} = \frac{\gamma_b}{32 G^2 g_{mc}^2 \kappa_c} \left(g_{mc}^2 \frac{(\kappa_c + \kappa_m)^ 2}{4} + \frac{\kappa_c^2 \kappa_m^2}{16} \right).
\end{equation}
The corresponding equivalent decay $\kappa_{\rm{imp, opt}}$ is always larger than both $\kappa_c$ and $\kappa_m$. 

\begin{figure}[h]
\includegraphics[width =  1.0 \columnwidth]{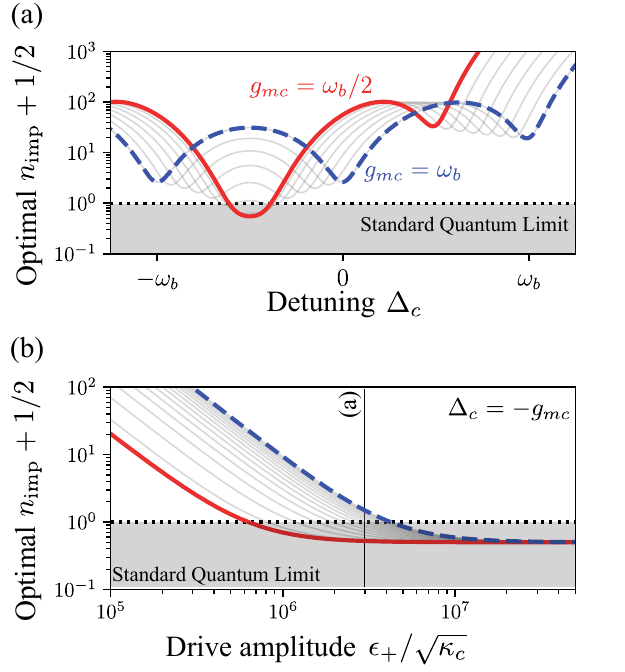}
\caption{\textbf{Number of added quanta due to imprecision noise for the optimal output quadrature.} (a) Number of imprecision quanta as a function of the cavity-drive detuning $\Delta_c$ and (b) as a function of the drive amplitude $\epsilon_+$.  Plots for magnon-microwave couplings ranging from $\omega_b/2$ (red curve) to $\omega_b$ (blue-dashed curve). In (a), we set the drive amplitude $\epsilon_+/\sqrt{\kappa_c} = 2 \times 10^6 $. The gray curves correspond to intermediate values of $g_{mc}$. The dotted line indicates the standard quantum limit for added noise $n_{\rm{imp}}^{\rm{SQL}}=1/2$. Plot for zero temperature and all parameters as in Table \ref{Table0}.} 
\label{Fig:nimp01}
\end{figure}

In Fig.~\ref{Fig:nimp01} (a), we show the optimal added quanta (offset by the vacuum fluctuations $1/2$) as a function of the drive-detuning $\Delta_c = \Delta_m$ for magnon-microwave couplings ranging from $\omega_b/2$ (triple resonance) to $\omega_b$ (dynamical backaction evasion). We observe in the curve the predicted minima of the added noise for detunings at the frequencies of the hybrid modes. The global minimum of added imprecision noise is obtained in the triple resonance scheme for a detuning set at the lower hybrid mode. In this case, the high-frequency tone is at the higher hybrid mode, while the lower-frequency tone is at the red mechanical sideband of the lower hybrid mode. The imprecision noise $n_{\rm{imp}}$ can be made arbitrarily small by increasing the driving amplitude. In such a perfect scenario, the added quanta to the measurement can go below the SQL, $n_{\rm{imp}}^{\rm{SQL}}=1/2$. We show the scaling of the added quanta (offset by the vacuum noise) as a function of power for tones centered around the lower hybrid mode in Fig.~\ref{Fig:nimp01} (b). In contrast with the single-tone case, shown in Fig.~\ref{Fig:AddednoiseStd}, the added noise in the BAE scheme decreases with the drive amplitude since the only source of added noise is the imprecision noise. Stronger magnon-microwave couplings require a stronger drive to beat the SQL, as we can see by comparing the results of the triple resonance configuration (continuous line) with the one for the dynamical backaction evasion configuration (dashed lines). We also notice that in the BAE scheme, it is possible to achieve a more precise measurement of the mechanics (less added noise) at weaker drive amplitudes in comparison with a single-tone drive. 

The analysis performed above relies on an RWA, which holds in the `good cavity' regime $\kappa_{c,m} <  \omega_b$ and for weak magnomechanical couplings. Analogously to standard optomechanical system \cite{clerk2008back}, we expect that our scheme can cope with corrections introduced by the RWA and still beat the quantum limit. The added noise also depends on the magnon/microwave bath occupancy and on the drive power \cite{SI}. For the parameters we considered here, the scheme is able to beat the SQL for magnon/microwave bath occupancies up to a few hundred quanta. For magnon/microwave modes with $\sim 10$ GHz frequencies, a negligible occupancy can be attained at a few hundreds of mK, which is routinely achieved in cavity magnonic experiments \cite{huebl2013high,lachance2019hybrid}.

\subsection*{Robustness of the backaction evasion scheme under imperfect tones separation}
\label{sec:robustness}

There are two requirements for BAE: the two tones applied to the microwave have to be separated by $2 \delta = 2 \omega_b$, and the amplitude of the tones has to be such that equation \eqref{eq:DriveAmplitudes} is satisfied. Nevertheless, it might be hard to achieve such requirements perfectly in practical setups. We consider here the effects of an imperfect tones separation in the added noise to the phonon measurement.

\begin{figure}[h]
\includegraphics[width = 1.0 \columnwidth]{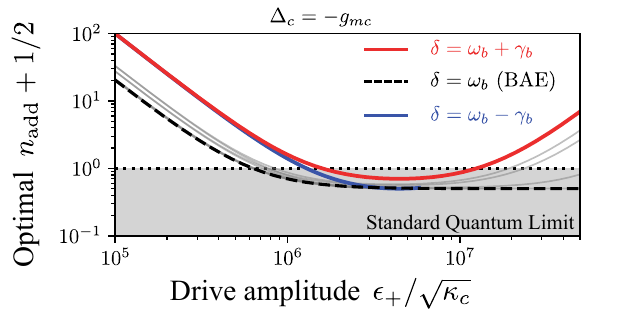}
\caption{\textbf{Added quanta to the measurement of the phonon mode for imperfect tone separation.} The total amount of added quanta (offset by the vacuum noise $1/2$) is shown as a function of the drive amplitude $\epsilon_+$ for several values of half of the tones frequency difference $\delta$ ranging from $\omega_b - \gamma_b$ (blue curve) to $\omega_b + \gamma_b$ (red curve). The magnon-microwave coupling is fixed at $g_{mc} = \omega_b/2$ (triple resonance scheme), and the drives are centered around the lower hybrid mode $\Delta_c = - g_{mc}$. The curve corresponding to the BAE scheme is indicated by a dashed line and the SQL by a dotted line. All parameters as in Table \ref{Table0}.  
} 
\label{Fig:PFinal}
\end{figure}

Deviations from the requirements for BAE change the measured phonon noise spectrum, adding backaction noise. The form of the measured output spectrum, in this case, resembles the one given in Eq.~\eqref{eq:NoiseSpectrumOneTone}, but with a modified backaction noise term, whose effect depends on both the mode splitting and the central tones frequency. We consider first the optimum point of the imprecision noise shown in Fig.~\ref{Fig:nimp01} (a), obtained for the triple resonance scheme and a central frequency with detuning $\Delta_c = -g_{mc}$. We show in Fig.~\ref{Fig:PFinal} the total added noise (imprecision plus backaction contributions) for different values of the tones frequency difference $\delta$. As the tone separation deviates from the BAE value, the added noise curve starts to become deformed at higher powers due to the backaction noise contribution. The imprecision noise contribution is also modified, getting stronger; nevertheless, we notice that even under such imperfect parameters, the measurement with a double tone can still beat the SQL at weaker drive amplitudes when compared with the single tone setup, compare with Fig.~\ref{Fig:nimp01}. For tones separation $< 2 \omega_b$, there is a drive amplitude range in which the measured phonon noise goes below the reference BAE value as a consequence of backaction squeezing of the mechanics and enhancement of the measurement gain. Correspondingly, the orthogonal quadrature is amplified. The system can also enter unstable regimes, at which the curves are abruptly ended. Similar results to those shown in Fig.~\ref{Fig:PFinal} are obtained in other cases, for instance, for imperfect drive amplitude balances, in correspondence with observations in electromechanical systems \cite{lecocq2015quantum}, and in agreement with recent proposals for generating microwave squeezing \cite{zhang2021generation,Qian_2024_strong}.

\begin{figure}[h]
\includegraphics[width = 1.0 \columnwidth]{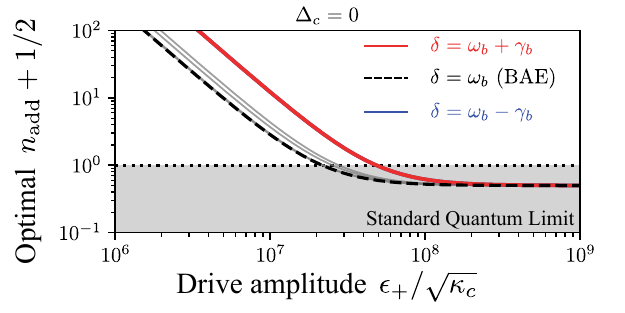}
\caption{\textbf{Added quanta to the measurement of the phonon mode for imperfect tone separation and zero detuning.} The total amount of added quanta (offset by the vacuum noise $1/2$) is shown as a function of the drive amplitude $\epsilon_+$ for several values of half of the tones frequency difference $\delta$ ranging from $\omega_b - \gamma_b$ (blue curve) to $\omega_b + \gamma_b$ (red curve). The magnon-microwave coupling is fixed at $g_{mc} = \omega_b/2$ (triple resonance scheme), and the drives are centered around the cavity/magnon frequency $\Delta_c = 0$. The curve corresponding to the BAE scheme is indicated by a dashed line and the SQL by a dotted line. All parameters as in Table \ref{Table0}. 
} 
\label{Fig:PFinal2}
\end{figure}

We can compare the robustness of the BAE scheme under a different choice parameters. As an example, we keep the hybrid mode splitting fixed, e.g. $g_{mc} = \omega_b/2$, and consider a different choice of the detuning $\Delta_c=0$. The total added noise for this case is shown in Fig.~\ref{Fig:PFinal2}. Even though this choice requires a stronger power to beat the SQL, we notice that there are no instabilities for the same range of added noise, and that, for the powers depicted here, the added noise curves show no apparent deformation due to backaction noise. Such facts suggest that this scheme is more robust to imperfect tone separation, even thought it does not provides the minimum of imprecision noise at a given drive power in the perfect BAE scheme. There is thus and interplay between robustness and optimum added noise to the measurement of the phonon mode that has to be considered for experimental implementations. 

Finally, results for imperfect drive amplitude balance exhibit a similar behavior to the ones shown here, pointing to the same conclusion about the interplay between robustness and minimum added noise.

\section*{Conclusion}
\label{sec:Concl}

To summarize, we have proposed and characterized a scheme for backaction evasion measurement of a mechanical quadrature tailored for cavity magnomechanical systems. Different schemes for backaction evasion are possible, and we have studied their robustness to imperfections of the BAE requirements, as well as the amount of noise added to the measurement via the output of the microwave mode. We have show that, at a given driven power, the minimum imprecision noise can be obtained for a system in triple resonance with drives centered around the lower frequency hybrid mode. In fact, in this case, state-of-the-art systems can go below the standard quantum limit at dilution fridge temperatures. Nevertheless, such a scheme is not the more robust to imperfections, exhibiting significant deviations from the BAE case which are less prominent in other configurations.

Our results provide a simple and flexible route for measuring phonons in cavity magnomechanics. For instance, such a quantum backaction evasion scheme can be used to perform quantum tomography of phonons and validate the creation of entangled and squeezed states proposed in the literature. A more complete model should include imperfect measurement efficiency as well as other intrinsic features of cavity magnomechanical systems, in particular, magnetic nonlinearities and the unavoidable coupling to high-order Walker modes \cite{bittencourt2023dynamical}. Both nonlinearities and the coupling to several magnon modes change dynamical backaction, and the same should be true for the quantum backaction studied here. Still, a double-drive BAE scheme should be possible, with small modifications due to the change in the magnon and phonon frequencies stemming from nonlinearities and the coupling to multiple magnon modes, an analysis that we postpone to future work.

\textit{Acknowledgements} V.A.S.V.B. and A.M. acknowledge financial support from the Contrat Triennal 2021-2023 Strasbourg Capitale Europeenne. C.A.P. acknowledges the support of the Natural Sciences and Engineering Research Council, Canada (NSERC). J.P.D. acknowledges funding support from the NSERC (Grant Nos.~RGPIN-2022-03078, and CREATE-495446-17); Alberta Innovates; and the Government of Canada through the NRC Quantum Sensors Program.


\clearpage
\onecolumngrid

\setcounter{equation}{0}
\setcounter{figure}{0}
\setcounter{table}{0}

\onecolumngrid

\section*{Supplementary Information}
\renewcommand{\theequation}{S.\arabic{equation}}

\section{Linearized Heisenberg-Langevin equations and noise correlations}
\renewcommand{\theequation}{S.\arabic{equation}}

We present briefly the linearized description of cavity magnomechanical systems that is used in the first section of the main text. We follow the same procedure used in the literature \cite{potts2020magnon,potts2021dynamical,potts2022dynamical,bittencourt2023dynamical}.

We start with the cavity magnomechanical Hamiltonian
\begin{equation}
\label{eq:simphamil}
\begin{aligned}
\frac{\hat{H}}{\hbar} &= \omega_c \hat{c}^\dagger \hat{c} + \omega_m \hat{m}^\dagger \hat{m} +\omega_b \hat{b}^\dagger \hat{b}+\frac{\hat{H}_{\rm{drive}}}{\hbar} \\
&+ g_{mc} \left( \hat{m}^\dagger \hat{c} + \hat{m} \hat{c}^\dagger \right) +g_{mb}^0 \hat{m}^\dagger \hat{m}\left(\hat{b}^\dagger +\hat{b} \right),
\end{aligned}
\end{equation}
with a single-tone drive term given by
\begin{equation}
\frac{\hat{H}_{\rm{drive}}}{\hbar} = i \sqrt{\kappa_e} \epsilon_d (\hat{c} e^{i \omega_d t} - \hat{c}^\dagger e^{-i \omega_d t}),
\end{equation}
with frequency $\omega_d$ and a real amplitude $\epsilon_d = \sqrt{\mathcal{P}/{\hbar \omega_d}}$, where $\mathcal{P}$ is the drive power. In correspondence with the main text, we will consider here the case $\omega_c = \omega_m$ (magnons and microwaves at resonance), and $\omega_d = \omega_{c,m}$ (drive at the cavity/magnon frequency).

Provided that the drive amplitude is strong and the magnomechanical coupling is small in comparison with the mode's frequencies and decays, we can adopt a linearized description for the system. In this case, for each mode $\alpha = c,m,b$  we write the annihilation operators as $\hat{\alpha} = \bar{\alpha}+\delta \hat{\alpha}$, where $\bar{\alpha}$ are coherent steady-state amplitudes. The Hamiltonian Eq.~\eqref{eq:simphamil} is then truncated up to second order on the fluctuations operators $\delta \hat{\alpha}$. From now on, we drop the $\delta$ indicating the fluctuations. The linearized magnomechanical Hamiltonian, in the frame rotating with the laser frequency $\omega_d$, can as
\begin{equation}
\label{eq:HamQuadra}
\begin{aligned}
\frac{\hat{H}}{\hbar} = &\omega_b \hat{b}^\dagger \hat{b} + g_{mc} (\hat{m}^\dagger \hat{c} + \hat{c}^\dagger \hat{m}) \\
&+ i \vert g_{mb} \vert (\hat{m}^\dagger - \hat{m})(\hat{b}^\dagger +\hat{b}),
\end{aligned}
\end{equation}
where $g_{mb} = \bar{m} g_{mb}^0$ is the linear magnomechanical coupling.

Through the paper, we model the open quantum dynamics of the system via Heisenberg-Langevin equations. For an annihilation operator $\hat{\alpha}$ for $\alpha = c,m,b$, we have
\begin{equation}
\partial_t \hat{\alpha} = \frac{i}{\hbar} [\hat{H}, \hat{\alpha}] - \frac{\kappa_{\alpha}}{2} \hat{\alpha} + \sqrt{\kappa_\alpha} \hat{\alpha}_{\rm{in}}(t).
\end{equation}

For the single-tone drive linearized Hamiltonian given in Eq.~(3), we have the following system of equations \cite{potts2021dynamical}
\begin{equation}
\begin{aligned}
\partial_t \hat{c} &= - \frac{\kappa_c}{2} \hat{c} - i g_{mc} \hat{m} + \sqrt{\kappa_c} \hat{c}_{\rm{in}}(t), \\
\partial_t \hat{m} &= - \frac{\kappa_m}{2} \hat{m} - i g_{mc} \hat{c} - ig_{mb}(\hat{b} + \hat{b}^\dagger) + \sqrt{\kappa_m} \hat{m}_{\rm{in}}(t), \\ 
\partial_t \hat{b} &= -(i \omega_b+ \frac{\gamma_b}{2}) \hat{b} - i g_{mb}\hat{m}^\dagger -i g_{mb}^* \hat{m} + \sqrt{\gamma_b} \hat{b}_{\rm{in}}(t).
\end{aligned}
\end{equation}
The noise terms included in the above equations are thermal and vacuum noises with correlations given by
\begin{equation}
\begin{aligned}
\langle \hat{\alpha}_{\rm{in}}(t) \hat{\alpha}_{\rm{in}}^\dagger(t^\prime) \rangle &= (n_\alpha + 1) \delta(t - t^\prime), \\
\langle \hat{\alpha}_{\rm{in}}^\dagger(t) \hat{\alpha}_{\rm{in}} (t^\prime) \rangle &= n_\alpha \delta(t - t^\prime),
\end{aligned}
\end{equation}
where $n_\alpha$ is the $\alpha$ mode bath occupancy. For magnons and microwaves we set $n_{c,m} = 0$.

The steady-state correlations of the system can be obtained by analyzing the equations of motion in frequency domain. For that, we use the Fourier transform:
\begin{equation}
\hat{\alpha}(t) = \int_{-\infty}^\infty \frac{d \omega}{2 \pi} e^{-i \omega t} \hat{\alpha}[\omega].
\end{equation}
Notice that
\begin{equation}
\hat{\alpha}^\dagger[\omega] = \int dt e^{i \omega t} \hat{\alpha}^\dagger(t) = (\hat{\alpha}[-\omega])^\dagger.
\end{equation}
The equations in frequency domain are
\begin{equation}
\label{eq:linhl}
\begin{aligned}
\chi_c^{-1}[\omega] \hat{c}[\omega] &= - i g_{mc} \hat{m}[\omega] + \sqrt{\kappa_c} \hat{c}_{\rm{in}}[\omega], \\
\chi_m^{-1}[\omega]  \hat{m} &= - i g_{mc} \hat{c}[\omega] - ig_{mb}(\hat{b}[\omega] + \hat{b}^\dagger[\omega]) + \sqrt{\kappa_m} \hat{m}_{\rm{in}}[\omega], \\ 
\chi_b^{-1}[\omega]  \hat{b}[\omega] &= - i g_{mb}\hat{m}^\dagger[\omega] -i g_{mb}^* \hat{m}[\omega] + \sqrt{\gamma_b} \hat{b}_{\rm{in}}[\omega],
\end{aligned}
\end{equation}
where the magnon and microwave susceptibilities are $\chi_{c,m}^{-1}[\omega] = -i \omega+\kappa_{c,m}/2$, and the phonon susceptibility is $\chi_{b}^{-1}[\omega] = -i (\omega-\omega_b)+\gamma_{b}/2$. In frequency domain, the noise correlations read
\begin{equation}
\begin{aligned}
\langle \hat{\alpha}_{\rm{in}}[\omega] \hat{\alpha}_{\rm{in}}^\dagger[\omega^\prime] \rangle &= 2 \pi (n_\alpha + 1) \delta[\omega - \omega^\prime], \\
\langle \hat{\alpha}_{\rm{in}}^\dagger[\omega] \hat{\alpha}_{\rm{in}} [\omega^\prime] \rangle &= 2 \pi n_\alpha \delta[\omega - \omega^\prime].
\end{aligned}
\end{equation}

By solving the linear set of Eqs.~\eqref{eq:linhl}, we obtain the solution for the phonon displacement quadrature given in Eq. (7) of the main text. In that equation $\hat{f}_{\rm{BA}}[\omega]$ is given by
\begin{equation}
\hat{f}_{\rm{BA}}[\omega] = g_{mb}^* \hat{\xi}_{\rm{BA}}[\omega] + g_{mb} \hat{\xi}_{\rm{BA}}^ \dagger[\omega],
\end{equation}
where
\begin{equation}
\begin{aligned}
\label{eq:xiBA}
\hat{\xi}_{\rm{BA}}[\omega] = \Xi_m [\omega] (\sqrt{\kappa_m} \hat{m}_{\rm{in}} [\omega]   - i g_{mc} \sqrt{\kappa_c} \chi_c [\omega] \hat{c}_{\rm{in}}  [\omega]).
\end{aligned}
\end{equation}
In the above equation $\Xi_m[\omega]$ is a modified magnon susceptibility given by
\begin{equation}
\Xi_m^ {-1}[\omega] = \chi_m^{-1}[\omega] +g_{mc}^2 \chi_c [\omega].
\end{equation}

\section{Output microwave spectrum for the single-tone drive}

We obtain the output spectrum from the standard input-output relation for the microwave mode
\begin{equation}
\hat{c}_{\rm{out}}[\omega] = \hat{c}_{\rm{in}}[\omega] - \sqrt{\kappa_c} \hat{c}[\omega].
\end{equation}
The microwave mode operator $\hat{c}[\omega]$ can be obtained by solving the linear Heisenberg-Langevin equations given in the last section. Defining
\begin{equation}
\hat{x}^{(0)}_b[\omega] = \sqrt{\frac{\gamma_b}{2}} \left( \chi_b[\omega] \hat{b}_{\rm{in}}[\omega] +  \chi_b^*[\omega] \hat{b}_{\rm{in}}^\dagger[\omega] \right) = \sqrt{\gamma_b} \hat{x}_{b,\rm{in}}[\omega],
\end{equation}
where we have defined 
\begin{equation}
\label{eq:xbin}
\hat{x}_{b,\rm{in}}[\omega] =\frac{\chi_b[\omega] \hat{b}_{\rm{in}}[\omega] +  \chi_b^*[-\omega] \hat{b}_{\rm{in}}^\dagger[\omega]}{\sqrt{2}}.
\end{equation}
We have the following expression for $\hat{c}_{\rm{out}}[\omega]$:
\begin{equation}
\begin{aligned}
\hat{c}_{\rm{out}}[\omega] &=  \sqrt{2\kappa_c} g_{mc} g_{mb} \chi_c[\omega] \Xi_m[\omega] \hat{x}_{b}^{(0)}[\omega] - i \sqrt{\kappa_c} g_{mc} g_{mb} \chi_c[\omega] \Xi_m[\omega]\mathcal{X}_{b}[\omega] \hat{f}_{\rm{BA}}[\omega] \\
&+ i g_{mc} \chi_c{\omega} \Xi_m[\omega] \sqrt{\kappa_c \kappa_m} \hat{m}_{\rm{in}}[\omega] + \left(1- \kappa_c  \frac{\chi_c[\omega] \Xi_m[\omega]}{\chi_m[\omega]} \right) \hat{c}_{\rm{in}}[\omega].
\end{aligned}
\end{equation}
All the quantities appearing in the above expression were defined in the previous section. We should notice that the in the left hand side of the above equation: the first term will correspond to the signal of the mechanical noise spectrum; the second term will contribute to the backaction noise, while the third term; the third and fourth terms will be the imprecision noise contribution, which depends on the magnon-microwave hybridization. We define then
\begin{equation}
\hat{\xi}_{\rm{imp}}[\omega] = i g_{mc} \chi_c[\omega] \Xi_m[\omega] \sqrt{\kappa_c \kappa_m} \hat{m}_{\rm{in}}[\omega] + \left(1- \kappa_c  \frac{\chi_c[\omega] \Xi_m[\omega]}{\chi_m[\omega]} \right) \hat{c}_{\rm{in}}[\omega], 
\end{equation}
and the following coefficients
\begin{equation}
\label{eqs:coeffstone}
\begin{aligned}
\mathcal{C}_x[\omega] &=   \sqrt{2\kappa_c} g_{mc} g_{mb} \chi_c[\omega] \Xi_m[\omega], \\
\mathcal{C}_{\rm{BA}}[\omega] &=  - i \sqrt{\kappa_c} g_{mc} g_{mb} \chi_c[\omega] \Xi_m[\omega]\mathcal{X}_{b}[\omega].
\end{aligned}
\end{equation}
Correlations of the noise operators $\hat{f}_{\rm{BA}}[\omega]$ and $\hat{\xi}_{\rm{imp}}[\omega]$ can be obtained using the correlations for the input noises $\hat{m}_{\rm{in}}[\omega]$ and $\hat{c}_{\rm{in}}[\omega]$ given in the previous section.

A quadrature of the output microwave mode is then given by
\begin{equation}
\begin{aligned}
\hat{x}_{{\rm{out}}, \theta }[\omega] &= \frac{\hat{c}_{\rm{out}}[\omega] e^{-i \theta} +\hat{c}_{\rm{out}}^\dagger[\omega] e^{i \theta}}{\sqrt{2}} \\
&= \mathcal{A}_{x, \theta}[\omega] \hat{x}^{(0)}_b[\omega] + \mathcal{A}_{{\rm{BA}},\theta}[\omega] \hat{f}_{\rm{BA}}[\omega] + \hat{x}_{{\rm{imp}},\theta }[\omega],
\end{aligned}
\end{equation}
where the coefficients $\mathcal{A}_{x,\theta}$ and $\mathcal{A}_{\rm{BA},\theta}$ are obtained directly from Eqs.~\eqref{eqs:coeffstone}. The output noise spectrum is then given by
\begin{equation}
\begin{aligned}
S_{\theta \theta}[\omega] &= \int_{-\infty}^{\infty} \frac{d\omega^\prime}{2 \pi} \langle \hat{x}_{{\rm{out}}, \theta }[\omega] \hat{x}_{{\rm{out}}, \theta }[\omega^\prime] \rangle, \\
&= \vert \mathcal{G}_{x, \theta} [\omega] \vert^2 S_{xx}^{(0)}[\omega] + S_{\rm{BA}, \theta}[\omega] + S_{\rm{imp}, \theta}[\omega] + S_{\rm{c}, \theta}[\omega].
\end{aligned}
\end{equation}
The correlation part $S_{\rm{c}, \theta}$ contains expectation values involves $\langle \hat{f}_{\rm{BA}}[\omega] \hat{x}_{\rm{imp},\theta}[\omega] \rangle$, which can be shown to be anti-symmetric in frequency. Since we are interested in symmetrized quantities, we do not show here the expression for $S_{\rm{c}, \theta}$. Otherwise, the other terms appearing in the output noise spectrum are given by
\begin{equation}
\begin{aligned}
\vert \mathcal{G}_{x, \theta} \vert^2 &= \frac{\vert \mathcal{A}_{x,\theta} \vert^ 2}{2}, \\
S_{\rm{BA}, \theta} &= \vert g_{mb} \vert^2 \vert \mathcal{A}_{\rm{BA}, \theta} [\omega]  \vert ^2 \Big[ \vert \Xi_m[\omega] \vert^2  \kappa_m (n_m+1) + g_{mc}^2 \vert^2 \vert \Xi_m[\omega] \vert^2 \vert \chi_c[\omega] \vert^2 \kappa_c (n_c +1) \\
&\quad \quad \quad \quad +  \vert \Xi_m[-\omega] \vert^2  \kappa_m  n_m + g_{mc}^2  \vert \Xi_m[-\omega] \vert^2 \vert \chi_c[-\omega] \vert^2 \kappa_c n_c \Big], \\
S_{\rm{imp}, \theta} &= \frac{1}{2} \Big[  g_{mc}^2 \kappa_c \kappa_m \vert \chi_c[\omega] \vert^2  \vert \Xi_m[\omega] \vert^2 (n_m +1) +  \vert 1- \kappa_c  \frac{\chi_c[\omega] \Xi_m[\omega]}{\chi_m[\omega]} \vert^2 (n_c+1) \\
&\quad \quad \quad \quad +  \quad g_{mc}^2 \kappa_c \kappa_m \vert \chi_c[-\omega] \vert^2  \vert \Xi_m[-\omega] \vert^2 n_m +  \vert 1- \kappa_c  \frac{\chi_c[-\omega] \Xi_m[-\omega]}{\chi_m[-\omega]} \vert^2 n_c \Big],
\end{aligned}
\end{equation}
and the mechanical spectrum is given in Eq.~(5). The parameters $n_{m,c}$ are the magnon and microwave baths occupancies, which we set to zero.

\section{Linearization and Heisenberg-Langevin equations for the two-tone drive}

The cavity magnomechanics Hamiltonian with the two-tone drive
\begin{equation}
\label{eq:FullHamiltonian2tones}
\begin{aligned}
\frac{\hat{H}}{\hbar} &= \omega_c \hat{c}^\dagger \hat{c} + \omega_m \hat{m}^\dagger \hat{m} +\omega_b \hat{b}^\dagger \hat{b} \\
&+ g_{mc} \left( \hat{m}^\dagger \hat{c} + \hat{m} \hat{c}^\dagger \right) +g_{mb}^0 \hat{m}^\dagger \hat{m}\left(\hat{b}^\dagger +\hat{b} \right)\\
 &+i \sqrt{\kappa_e} (\epsilon_{-} e^{- i \delta t} + \epsilon_{+} e^{i \delta t})e^{i \omega_d t} \hat{c} + \rm{H.c.}.
\end{aligned}
\end{equation}
To obtain the semiclassical steady-state, we move to an interacting frame rotating with $\omega_d \hat{c}^\dagger \hat{c}+ \omega_d \hat{m}^\dagger \hat{m}$ and make the ansatz $\hat{c} = \hat{c}_0 + \hat{c}_{+} e^{-i \delta t} + \hat{c}_{-} e^{i \delta t}$ and $\hat{m} = \hat{m}_0 + \hat{m}_{+} e^{-i \delta t} + \hat{m}_{-} e^{i \delta t}$. Within a mean field approximation, the equations for the expectation values of each bosonic operator reads
\begin{equation}
    \begin{aligned}
        \dot{c}_0 &= \left( i \Delta_c - \frac{\kappa}{2} \right) c_0- i g_{mc} m_0, \\
        \dot{c}_{\pm } &= \left( i (\Delta_c \pm \delta) - \frac{\kappa}{2} \right) c_{\pm }- i g_{mc} m_{\pm } - \sqrt{\kappa_e} \epsilon_{\pm}, \\
        \dot{m}_0 &= \left( i \Delta_m - \frac{\gamma}{2} \right) m_0 - i g_{mc} c_0 - i g_{mb}^0 (m_{-} b + m_{+} b^*), \\
        \dot{m}_{+}  &= \left( i (\Delta_m + \delta) - \frac{\gamma}{2} \right) m_{+}  - i g_{mc} c_{+} - i g_{mb}^{0} m_0 b, \\
        \dot{m}_{-}  &= \left( i (\Delta_m - \delta) - \frac{\gamma}{2} \right) m_{-}  - i g_{mc} c_{-} - i g_{mb}^{0} m_0 b^*, \\
         \dot{b} &= \left( i \Delta_b - \frac{\Gamma}{2} \right) b  - i g_{mb}^0 (m_0^* m_{+} + m_{-}^* m_0).
    \end{aligned}
\end{equation}
We have discarded all terms $\propto e^{\pm 2 i \delta t}$. The steady-state for the operators $\bar{ o}_i$  are obtained by setting all time derivatives to zero and solving the system of equations, which gives
\begin{equation}
\label{eq:steadystate2tones}
\begin{aligned}
    \bar{c}_0 &= \bar{m}_0 = 0, \\
    \bar{c}_{\pm} &= \frac{\sqrt{\kappa_e} \epsilon_{\pm}  \left( i (\Delta_m \pm \delta) - \frac{\kappa_m}{2} \right) }{ \left( i (\Delta_m \pm \delta) - \frac{\kappa_m}{2} \right) \left( i (\Delta_c \pm \delta) - \frac{\kappa_c}{2} \right)  + g_{mc}^2 }, \\
    \bar{m}_{\pm} &= \frac{i g_{mc} \sqrt{\kappa_e} \epsilon_{\pm}  }{ \left( i (\Delta_m \pm \delta) - \frac{\kappa_m}{2} \right) \left( i (\Delta_c \pm \delta) - \frac{\kappa_c}{2} \right)  + g_{mc}^2 }.
\end{aligned}
\end{equation}
In doing so, we have assumed that the bare magnomechanical coupling $g_{mc}^0$ is small enough to be ignored in the dynamics of the magnon and microwave amplitudes.

We linearize the two-tone drive Hamiltonian by considering fluctuations around the steady-state given by Eqs. \eqref{eq:steadystate2tones}: $\hat{\alpha} = \delta \hat{\alpha} + \bar{\alpha}_{+} e^{-i \delta t}+ \bar{\alpha}_{- } e^{i \delta t} $ for $\alpha = c, m$. We drop the deltas indicating the fluctuations and keep only the quadratic terms of the Hamiltonian. In an interacting frame with respect to $\omega_b \hat{b}^\dagger \hat{b}$, the Hamiltonian reads
\begin{equation}
\label{eq:LinHamil01}
\begin{aligned}
\frac{\hat{H}_L^{(2)}}{\hbar}  &= -\Delta_c \hat{c}^\dagger \hat{c} -\Delta_m \hat{m}^\dagger \hat{m} + g_{mc} \left( \hat{m}^\dagger \hat{c} + \hat{m} \hat{c}^\dagger \right) \\
&+ e^{i \delta t} \left( g_{-} \hat{m}^\dagger + g_{+}^* \hat{m} \right) \left( \hat{b} e^{-i \Omega_b t} + \hat{b}^\dagger e^{i \Omega_b t} \right) \\
&+ e^{-i \delta t} \left( g_{+} \hat{m}^\dagger + g_{-}^* \hat{m} \right) \left( \hat{b} e^{-i \Omega_b t} + \hat{b}^\dagger e^{i \Omega_b t} \right),
\end{aligned}
\end{equation}
where we have defined $g_{\pm }= g_{mb}^0 \bar{m}_{\pm}$.  Since $\delta$ is positive, the interacting terms are resonant provided that $\delta = \omega_b$, in other words, if the two tones are separated by twice the phonon frequency. With this choice, the QND Hamiltonian for a phonon quadrature can be obtained by an appropriate choice of the magnomechanical coupling rates $g_{\pm} = \vert g_{\pm} \vert e^{i \phi_\pm}$. If coupling rates have the same modulus,
\begin{equation}
\label{eq:requirement}
\vert g_{-} \vert = \vert g_{+} \vert = G,
\end{equation}
we get after some algebraic manipulations
\begin{equation}
\label{eq:qndhamiltonian}
\begin{aligned}
\frac{\hat{H}_{\rm{QND}}}{\hbar}  &= -\Delta_c \hat{c}^\dagger \hat{c} -\Delta_m \hat{m}^\dagger \hat{m} + g_{mc} \left( \hat{m}^\dagger \hat{c} + \hat{m} \hat{c}^\dagger \right) \\
&+  G \left( e^{i \varphi} \hat{m}^\dagger + e^{-i \varphi} \hat{m} \right) \left( e^{i \psi} \hat{b}^\dagger + e^{-i \psi} \hat{b} \right),
\end{aligned}
\end{equation}
where $\varphi = (\phi_+ + \phi_-)/2$ and $\psi =(\phi_+ - \phi_-)/2$. The equality given in Eq.~\eqref{eq:requirement} implies that  $ \vert \bar{m}_+ \vert = \vert \bar{m}_- \vert $ thus yielding the balance condition for the drive amplitudes
\begin{equation}
\frac{\vert \epsilon_{+} \vert}{\vert \epsilon_{- } \vert}= \frac{\vert \left( i (\Delta_m + \omega_b) - \frac{\kappa_m}{2} \right) \left( i (\Delta_c + \omega_b) - \frac{\kappa_c}{2} \right)  + g_{mc}^2 \vert }{\vert \left( i (\Delta_m - \omega_b) - \frac{\kappa_m}{2} \right) \left( i (\Delta_c - \omega_b) - \frac{\kappa_c}{2} \right)  + g_{mc}^2 \vert}.
\end{equation}
In the basis of the quadratures, Eq.~\eqref{eq:qndhamiltonian} is given by the main text Eq.~(22).

\section{Solution for the phonon operator in the linearized two-tone driving scheme}

We define $\delta_b = \omega_b - \delta$ and $\sigma_b = \omega_b+\delta$. In an interacting frame rotating with $\omega_b \hat{b}^\dagger \hat{b}$, we have the following equation for $\hat{m}[\omega]$ as a function of phonon operators
\begin{equation}
\begin{aligned}
 \hat{m}[\omega] = - i g_- \Xi[\omega] \left( \hat{b}[\omega - \delta_b] + \hat{b}^\dagger [\omega  + \sigma_b] \right) - i g_+ \Xi[\omega] (\hat{b}[\omega - \sigma_b] + \hat{b}^\dagger[\omega + \delta_b]) + \hat{\xi}_{\rm{BA}}[\omega],
\end{aligned}
\end{equation}
where the modified magnon susceptibility $\Xi[\omega]$ and the backaction noise $\hat{\xi}_{\rm{BA}}[\omega]$ have the same formula as in the single drive case. The equation for the phonon operator reads
\begin{equation}
\chi_b^{-1}[\omega] \hat{b}[\omega] = - i g_- \hat{m}^\dagger[\omega+\sigma_b] - i g_+ \hat{m}^\dagger[\omega+\delta_b] - i g_-^* \hat{m}[\omega + \delta_b] - i g_+^* \hat{m}[\omega + \sigma_b] +\sqrt{\gamma_b}\hat{b}_{\rm{in}}[\omega].
\end{equation}
Notice that in this frame $\chi_b[\omega] = (-i \omega + \gamma_b/2)^{-1}$.

Eliminating the magnon operator in favor of the phonon operator gives
\begin{equation}
\label{eq:fullb}
\begin{aligned}
\chi^{-1}_b[\omega] \hat{b}[\omega] &= i \Sigma_{b, {\rm{T}} }  [\omega] \hat{b}[\omega] + i \Sigma_{b, {\rm{T}}} [\omega] \hat{b}^\dagger [\omega + 2 \omega_b] + f_1[\omega] \hat{b}[\omega + 2 \delta] + f_1[\omega]\hat{b}^\dagger [\omega + 2 \sigma_b] \\
&+ f_2 [\omega] \hat{b}[\omega - 2 \delta] + f_2[\omega] \hat{b}^\dagger[\omega + 2 \delta_b] - i g_+ \hat{\xi}_{\rm{BA}}^\dagger[\omega + \delta_b] - i g_-^* \hat{\xi}_{\rm{BA}}[\omega + \delta_b] \\
&- i g_- \hat{\xi}_{\rm{BA}}^\dagger[\omega + \sigma_b] - i g_+^*\hat{\xi}_{\rm{BA}}[\omega +\sigma_b] +\sqrt{\gamma_b} \hat{b}_{\rm{in}} [\omega],
\end{aligned}
\end{equation}
In the above expression, the total phonon self-energy $\Sigma_{b, {\rm{T}}} [\omega]$ is given by
\begin{equation}
\Sigma_{b, {\rm{T}}} [\omega] = \Sigma_{b} [\omega] + \Sigma_{b,{\rm{CRT}} } [\omega], 
\end{equation}
where the co-rotating contribution $\Sigma_{b} [\omega]$ is given by
\begin{equation}
\label{eq:phononselfenergyfull}
\Sigma_{b} [\omega] = i \vert g_- \vert^2 \Xi_m[\omega + \delta_b] - i \vert g_+ \vert^2 \Xi_m^*[-\omega - \delta_b],
\end{equation}
which is similar to that obtained for the one-tone drive case, while the \textit{counter-rotating} contribution is
\begin{equation}
\Sigma_{b,{\rm{CRT}} } [\omega] = i \vert g_+ \vert^2 \Xi_m[\omega + \sigma_b] -i \vert g_- \vert^2 \Xi_m^*[-\omega - \sigma_b].
\end{equation}
The pre-factor functions $f_{1,2} [\omega]$ are given by
\begin{equation}
\label{eq:f1f2}
\begin{aligned}
f_1[\omega] &= g_- g_+^* \left( \Xi_m^*[- \omega - \sigma_b] - \Xi_m[ \omega + \sigma_b] \right), \\
f_2[\omega] &= g_-^* g_+ \left( \Xi_m^*[- \omega - \delta_b] - \Xi_m[ \omega + \delta_b] \right).
\end{aligned}
\end{equation}

\subsection{Mechanical noise spectrum in the BAE scheme}

The BAE scheme requires that $\delta_b = 0$ and $\vert g_+ \vert = \vert g_- \vert = G$. Using the same notation of the main text $g_\pm = G e^{\pm i  \phi_\pm}$, we obtain for the pre-factors
\begin{equation}
\begin{aligned}
\Sigma_{b, {\rm{T}}} [\omega] &= \Sigma_{b} [\omega] + \Sigma_{b,{\rm{CRT}} } [\omega],  \\
\Sigma_{b} [\omega] &= i G ^2 \left( \Xi_m[\omega] - \Xi_m^*[-\omega] \right), \\
\Sigma_{b,{\rm{CRT}} } [\omega] &= i  G^2 \left( \Xi_m[\omega + 2 \omega_b] - \Xi_m^*[-\omega - 2 \omega_b] \right), \\
f_1[\omega]&= -G^2 e^{-i(\phi_+ - \phi_-)} \left(  \Xi_m[\omega + 2 \omega_b] - \Xi_m^*[-\omega - 2 \omega_b] \right) =  i e^{-i(\phi_+ - \phi_-)} \Sigma_{b,{\rm{CRT}} } [\omega] \\
f_2 [\omega] &= - G^2 e^{i (\phi_+ - \phi_-)} \left(  \Xi_m[\omega] - \Xi_m^*[-\omega] \right) = i  e^{i (\phi_+ - \phi_-)} \Sigma_{b} [\omega].
\end{aligned}
\end{equation}
and the phonon operator is given by
\begin{equation}
\label{eqapp:fullb}
\begin{aligned}
\chi^{-1}_b[\omega] \hat{b}[\omega] &= i \Sigma_{b}  [\omega] \hat{b}[\omega] +i \Sigma_{b, \rm{CRT}}  [\omega] \hat{b}[\omega] + i \Sigma_{b, {\rm{T}}} [\omega] \hat{b}^\dagger [\omega + 2 \omega_b] \\
&+ i e^{-2 i\psi} \Sigma_{b,{\rm{CRT}} } [\omega] \left( \hat{b}[\omega + 2 \omega_b] + \hat{b}^\dagger [\omega + 4 \omega_b] \right) \\
&+ i  e^{2 i \psi} \Sigma_{b} [\omega] \left( \hat{b}^\dagger[\omega]+ \hat{b}[\omega - 2 \omega_b]   \right) \\
&- i G e^{i (\psi + \varphi)} \hat{\xi}_{\rm{BA}}^\dagger[\omega ] - i G e^{i (\psi - \varphi)}\hat{\xi}_{\rm{BA}}[\omega] \\
&- i G e^{-i (\psi - \varphi)} \hat{\xi}_{\rm{BA}}^\dagger[\omega + 2 \omega_b] - i G e^{-i (\psi + \varphi)} \hat{\xi}_{\rm{BA}}[\omega +2 \omega_b] +\sqrt{\gamma_b} \hat{b}_{\rm{in}} [\omega],
\end{aligned}
\end{equation}
where we have explicitly separated rotating and counter-rotating contributions for the self-energy in the first term and we have used the definitions. We have defined the phases $\psi$ and $\varphi$ as
\begin{equation}
\begin{aligned}
\psi = \frac{\varphi_+ - \varphi_-}{2}, \\
\varphi = \frac{\phi_+ + \phi_-}{2}.
\end{aligned}
\end{equation}

After some algebraic manipulations, from Eq. \eqref{eqapp:fullb} we obtain the following equation for the quadrature $\hat{x}_{b, \psi}$
\begin{equation}
\label{eq:positionopfull}
\begin{aligned}
\sqrt{2} \chi^{-1}_b[\omega] \hat{x}_{b, \psi}[\omega] &= \sqrt{2 \gamma_b} (\hat{b}_{\rm{in}}[\omega] e^{- i \psi} + \hat{b}^\dagger_{\rm{in}}[\omega] e^{ i \psi}) \\
&+i \Sigma_{b, \rm{CRT}}  [\omega] e^{-i \psi} \hat{b}[\omega] - i \Sigma_{b, \rm{CRT}}^*  [-\omega] e^{i \psi} \hat{b}^\dagger[\omega] \\
&+ i e^{-3 i\psi} \Sigma_{b,{\rm{CRT}} } [\omega] \left( \hat{b}[\omega + 2 \omega_b] + \hat{b}^\dagger [\omega + 4 \omega_b] \right) - i e^{3 i \psi} \Sigma_{b,{\rm{CRT}} }^* [-\omega] \left( \hat{b}^\dagger[\omega - 2 \omega_b] + \hat{b} [\omega - 4 \omega_b] \right) \\
& + i   \Sigma_{b} [\omega] \left( e^{ i \psi} \hat{b}[\omega - 2 \omega_b] -  e^{- i \psi} \hat{b}^\dagger[\omega + 2 \omega_b] \right) \\
& - i G e^{-i (2 \psi - \varphi)} \hat{\xi}_{\rm{BA}}^\dagger[\omega + 2 \omega_b] + i G e^{i (2 \psi - \varphi)} \hat{\xi}_{\rm{BA}}[\omega - 2 \omega_b] \\
&-i G e^{-i (2 \psi + \varphi)} \hat{\xi}_{\rm{BA}}[\omega +2 \omega_b] + i G e^{i (2 \psi + \varphi)} \hat{\xi}_{\rm{BA}}^\dagger[\omega -2 \omega_b].
\end{aligned}
\end{equation}
All the terms on the right-hand side of the above equation, besides the first one, are counter-rotating terms, either because they are multiplied by $\Sigma_{b, {\rm{CRT}}}$ or because they are evaluated at twice a phonon frequency apart from the frequency where the equation is evaluated. In general, one can  write an infinite set of linear equation for the quadrature and its canonical conjugated $\hat{p}_{b, \psi}$,  evaluated at frequencies $\omega \pm 2 n \omega_b$, where $n$ is a positive integer. 

Alternatively, we can perform the rotating wave approximation in Eq.~\eqref{eqapp:fullb}, which yields the following set of equation for $e^{- i \psi} \hat{b}[\omega]$ and  $e^{i \psi} \hat{b}^\dagger[\omega]$ 
\begin{equation}
\begin{aligned}
\chi_b^{-1}[\omega] e^{- i \psi} \hat{b}[\omega] &= i e^{-i \psi} \Sigma_b[\omega] \hat{b}[\omega] + i e^{i \psi} \Sigma_b[\omega] \hat{b}^\dagger[\omega] - i G e^{i \varphi} \hat{\xi}_{\rm{BA}}^\dagger[\omega]- i G e^{-i \varphi} \hat{\xi}_{\rm{BA}}[\omega] + \sqrt{\gamma_b} e^{- i  \psi} \hat{b}_{\rm{in}}[\omega], \\
\chi_b^{-1}[\omega] e^{ i \psi} \hat{b}^\dagger[\omega] &= -i e^{i \psi} \Sigma_b[\omega] \hat{b}^\dagger[\omega] - i e^{-i \psi} \Sigma_b[\omega] \hat{b}[\omega] + i G e^{-i \varphi} \hat{\xi}_{\rm{BA}}[\omega]+ i G e^{i \varphi} \hat{\xi}_{\rm{BA}}^\dagger[\omega] + \sqrt{\gamma_b} e^{ i  \psi} \hat{b}_{\rm{in}}^\dagger[\omega].
\end{aligned}
\end{equation}

We then define the following noise term
\begin{equation}
\begin{aligned}
\hat{x}_{b,\psi,\rm{in}}[\omega] &= \chi_b[\omega] \frac{e^{i \psi} \hat{b}_{\rm{in}}^\dagger[\omega]  + e^{-i \psi} \hat{b}_{\rm{in}}[\omega]}
{\sqrt{2}}. \\
\hat{p}_{b,\psi,\rm{in}}[\omega] &= i \chi_b[\omega] \frac{e^{i \psi} \hat{b}_{\rm{in}}^\dagger[\omega]  - e^{-i \psi} \hat{b}_{\rm{in}}[\omega]}
{\sqrt{2}}, \\
\hat{\xi}_{{\rm{BA}}, x_{\varphi}}[\omega] &= \frac{e^{i \varphi} \hat{\xi}_{\rm{BA}}^\dagger[\omega] + e^{-i \varphi} \hat{\xi}_{\rm{BA}}[\omega]}{\sqrt{2}},
\end{aligned}
\end{equation}
such that the equations for the phonon quadratures read
\begin{equation}
\begin{aligned}
\hat{x}_{b, \psi}[\omega] &= \sqrt{\gamma_b} \hat{x}_{b, \psi, \rm{in}}[\omega],\\
\hat{p}_{b, \psi}[\omega] &=  \sqrt{\gamma_b} \hat{p}_{b,\psi,\rm{in}}[\omega]  - 2 G \chi_b[\omega] \hat{\xi}_{{\rm{BA}}, x_{\varphi}}[\omega] + 2  \chi_b[\omega] \Sigma_b[\omega] \hat{x}_{b, \psi}[\omega].
\end{aligned}
\end{equation}
Substituting the first equation into the second gives Eqs.~(14) in the main text.

\textit{If in addition to the other requirements} we further impose that both drives are centered around the magnon/microwave frequency, i.e., if the drive-detuning is zero, then as in the one-tone drive case $\Sigma_b[\omega] = \Sigma_{b, \rm{CRT}} [\omega] = 0$. In this particular case, Eq.~\eqref{eq:positionopfull}, which includes the counter-rotating terms, yields:
\begin{equation}
\begin{aligned}
\sqrt{2} \chi^{-1}_b[\omega] \hat{x}_{b, \psi}[\omega] &= \sqrt{2 \gamma_b}(\hat{b}_{\rm{in}}[\omega] e^{- i \psi} + \hat{b}^\dagger_{\rm{in}}[\omega] e^{ i \psi}) \\
& - i G e^{-i (2 \psi - \varphi)} \hat{\xi}_{\rm{BA}}^\dagger[\omega + 2 \omega_b] + i G e^{i (2 \psi - \varphi)} \hat{\xi}_{\rm{BA}}[\omega - 2 \omega_b] \\
&-i G e^{-i (2 \psi + \varphi)} \hat{\xi}_{\rm{BA}}[\omega +2 \omega_b] + i G e^{i (2 \psi + \varphi)} \hat{\xi}_{\rm{BA}}^\dagger[\omega -2 \omega_b].
\end{aligned}
\end{equation}
Since, in this situation, microwaves and magnons are at resonance, $n_m = n_c$. The noise spectrum of such quadrature is then given by
\begin{equation}
S_{xx, \psi}[\omega] = \int \frac{d \omega^\prime}{2 \pi} \langle \hat{x}_{b, \psi}[\omega] \hat{x}_{b, \psi}[\omega^\prime] \rangle = S_{xx, \psi}^{(0)}[\omega] + S_{xx, \psi}^{(\rm{CRT})}[\omega],
\end{equation}
where
\begin{equation}
S_{xx, \psi}^{(0)}[\omega] = \gamma_b \vert \chi_b [\omega] \vert^2 \left( n_b + \frac{1}{2}\right) = \frac{\gamma_b \left( n_b + \frac{1}{2}\right)}{\omega^2 + \frac{\gamma_b^2}{4}}, 
\end{equation}
is the uncoupled noise spectrum, and
\begin{equation}
\begin{aligned}
S_{xx, \psi}^{(\rm{CRT})}[\omega] =  \vert \chi_b [\omega] \vert^2 G^2  \left( n_{m} + \frac{1}{2}\right) &\Bigg[ \mathcal{A}[\omega - 2 \omega_b] \left(1 - \frac{\chi_b^*[\omega - 4 \omega_b] e^{4 i \psi}}{\chi_b^*[\omega]} \right) \\ &\quad + \mathcal{A}[\omega + 2 \omega_b] \left(1 - \frac{\chi_b^*[\omega + 4 \omega_b] e^{-4 i \psi}}{\chi_b^*[\omega]} \right)  \Bigg],
\end{aligned}
\end{equation}
where
\begin{equation}
\mathcal{A}[\omega] = \vert \Xi_m[\omega] \vert^2 (\kappa_m + g_{mc}^2 \kappa_c \vert \chi_c [\omega] \vert^2).
\end{equation}
Notice that since the drive detuning $\Delta_c = \Delta_m = 0$ then $\mathcal{A}[\omega] =\mathcal{A}[-\omega]$. Furthermore $\chi_b [\omega] = \chi_b^*[-\omega]$.

\begin{figure}[h]
\includegraphics[width =0.5\columnwidth]{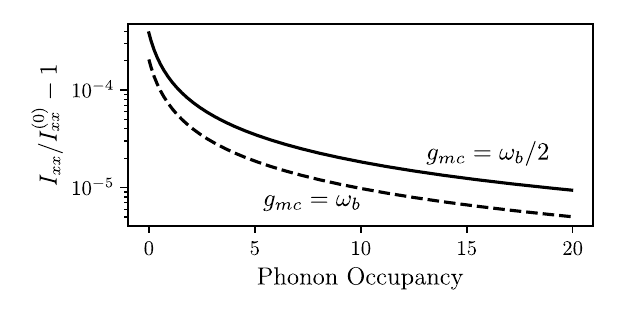}
\caption{Ratio between the integral of the noise spectrum of the BAE quadrature, including the counter-rotating terms and its value without counter-rotating terms, for both the triple resonance and the dynamical backaction evasion schemes. Parameters as in Table 1 of the main text.} 
\label{Fig:Appendix03}
\end{figure}

In Fig.~\ref{Fig:Appendix03}, we compute the effects of the counter-rotating terms in the QND quadrature noise spectrum by evaluating its integral and comparing it with the value for a resonator in thermal equilibrium. We notice that, for the chosen parameters, in the worst case, the corrections introduced by the CRTs do not exceed $~4\, 10^{-4} I_{xx}^ {(0)}$, which can be shown to be around two orders of magnitude smaller than the same quantity evaluated for the single drive case shown. In this two-tone case, the CRTs would add a negligible amount of noise when compared to the noise added by measurement backaction in the single-tone case at zero detuning. We also notice that errors introduced by the CRTs are typically an order of magnitude smaller than the error introduced by imperfections on BAE conditions, see section IIIB.

In the case where the detunings do not vanish, the analysis of the effects of counter-rotating terms is more involving. The position quadrature operator is given by Eq.~\eqref{eq:positionopfull}, which depends on the phonon operators  $
\hat{b}^{(\dagger)}[\omega \pm 2 \omega_b]$ and $\hat{b}^{(\dagger)}[
\omega \pm 4 \omega_b]$. As mentioned above, one then has to solve an infinite system of linear equations, which can be truncated to a given order. We postpone such an analysis to a future work, but we expect that, as in the case $\Delta_m = \Delta_c = 0$, the corrections due to discarded counter-rotating terms are small. 

\section{Momentum noise spectrum}

From the Eqs. (14) of the main text, we get the noise spectrum of the momentum quadrature
\begin{equation}
S_{pp}[\omega] = S^{(0)}[\omega] +  S_{pp}^{(\rm{D})}[\omega] + S_{pp}^{(\rm{BA})}[\omega] + S_{pp}^{(\rm{DBA})}[\omega].
\end{equation}
The terms $S^{(0)}[\omega]$ has the same form as the uncoupled phonon noise spectrum. The backaction contribution is given by
\begin{equation}
\label{eq:SpppBA}
\begin{aligned}
S_{pp}^{(\rm{BA})}[\omega] &= 4 \frac{G^2 \vert \Xi_m [\omega] \vert^2}{\omega^2 + \frac{\gamma_b^2}{4}} \Big[ \kappa_m \left(n_m +\frac{1}{2} \right) + g_{mc}^2 \kappa_c \vert \chi_c [\omega] \vert^2 \left( n_c + \frac{1}{2} \right)\Big],
\end{aligned}
\end{equation}
The additional contributions are $S_{pp}^{(\rm{D })}[\omega]$, a term that comes from the phonon noise and depends on the phonon self-energy given by
\begin{equation}
S_{pp}^{(\rm{D })}[\omega]=  \frac{4 \vert \Sigma_b [\omega] \vert^ 2}{\omega^2 + \frac{\gamma_b^2}{4}} \left(n_b + \frac{1}{2} \right),
\end{equation}
while the second backaction noise contribution $S_{pp}^{(\rm{DBA})}[\omega]$ is given by
\begin{equation}
S_{pp}^{(\rm{DBA})}[\omega] = - \frac{2 \sqrt{\gamma_b} {\rm{Im}}[\chi_b[\omega] \Sigma_b[\omega]]}{\omega^2 + \frac{\gamma_b^2}{4}}.
\end{equation}

To compute the effects of backaction noise in the non-BAE quadrature of the phonon mode, we compute the integral of the noise spectrum
\begin{equation}
I_{pp} = \int \frac{d\omega}{2 \pi} S_{pp}[\omega].
\end{equation}
We take the value for an uncoupled oscillator of the same frequency at thermal equilibrium value as a reference
\begin{equation}
I_{pp}^{(0)} = \int \frac{d\omega}{2 \pi} S^{(0)}[\omega] = n_b + \frac{1}{2}.
\end{equation}

\begin{figure}[t]
\includegraphics[width =0.4\columnwidth]{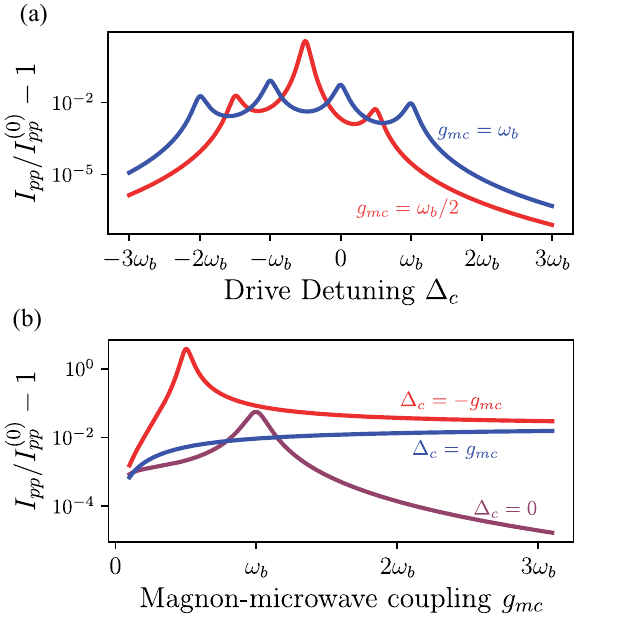}
\caption{Ratio between the uncertainty in the non-QND quadrature $\hat{p}_{\psi}$ and the reference value for an uncoupled oscillator in thermal equilibrium (a) as a function of the detuning between the central frequency of the two-tones and the microwave frequency for the triple resonant scheme and the dynamical backaction evasion scheme, and (b) as a function of the magnon-microwave coupling for different values of $\Delta_c$. Plots for zero temperature and all parameters as in Table 1.} 
\label{Fig:IntegralPP}
\end{figure}

In Fig.~\ref{Fig:IntegralPP}, we show the ratio $I_{pp}/I_{pp}^{0}$ as a function of the detuning $\Delta_c$, and as a function of the magnon-microwave coupling. Different magnomechanics schemes yield different peaks in the added noise for the non-BAE quadrature. For the triple resonance scheme, we notice maxima of $I_{pp}$ for detunings $\Delta_c = (-3 \omega_b/2, -\omega_b/2, \omega_b/2)$, while for the dynamical backaction evasion scheme the maxima are at $\Delta_c = (-2 \omega_b, -\omega_b, 0, \omega_b)$. We can understand the position of these peaks by considering the hybrid mode formation in each case. For example, in the case of the highest peak shown in Fig.~\ref{Fig:IntegralPP}(a), since $g_{mc} = \omega_b/2$, the magnon-microwave hybrid modes are located at $\omega_c \pm \omega_b/2$. The red sideband of the upper hybrid mode coincides with the frequency of the lower hybrid mode, and the blue sideband of the lower hybrid mode coincides with the frequency of the higher hybrid mode. When the detuning is set at $\Delta_c = -\omega_b/2$, the high-frequency drive is at the frequency of the upper hybrid mode, driving it and the blue sideband of the lower hybrid mode while the low-frequency drive is at the red sideband of the lower hybrid mode.

\section{Output microwave noise spectrum}

Here, we show the formulas for the output microwave noise spectrum that are not displayed in the main text. We consider exclusively the case where the BAE conditions are met. From the Heisenberg-Langevin equations, we obtain the following solution for the microwave operator as a function of the noises and the BAE phonon quadrature
\begin{equation}
\begin{aligned}
\chi_c^{-1}[\omega]\hat{c}[\omega] = - \sqrt{2} G g_{mc} \Xi_m[\omega] e^{i \varphi} \hat{x}_{b, \psi}[\omega] - i g_{mc} \hat{\xi}_{\rm{BA}}[\omega] + \sqrt{\kappa_c} \hat{c}_{\rm{in}}[\omega].
\end{aligned}
\end{equation}
Recall that the ``backaction`` noise term $\hat{\xi}_{\rm{BA}}[\omega]$ has a component $\hat{c}_{\rm{in}}[\omega]$, see Eq. \eqref{eq:xiBA}. We now consider the standard input-output relation for the reflection of the microwave mode and assume that the input noise is white, such that
\begin{equation}
\hat{c}_{\rm{out}}[\omega] = \hat{c}_{\rm{in}}[\omega] - \sqrt{\kappa_c} \hat{c} [\omega].
\end{equation}
We then compute the output quadrature
\begin{equation}
\begin{aligned}
\hat{x}_{\rm{out,\theta}}[\omega] &= \frac{e^{i \theta}\hat{c}^\dagger_{\rm{out}}[\omega] + e^{-i \theta}\hat{c}_{\rm{out}}[\omega]}{\sqrt{2}} \\
&= \mathcal{G}_{x, \theta} [\omega] \hat{x}_{b, \psi}[\omega] + \mathbb{A}_{m;\theta} [\omega] \hat{m}_{\rm{in}}[\omega] + \mathbb{A}_{c;\theta} [\omega] \hat{c}_{\rm{in}}[\omega] \\
&\quad +\mathbb{A}_{m;\theta}^* [-\omega] \hat{m}_{\rm{in}}^\dagger [\omega] + \mathbb{A}_{c;\theta}^* [-\omega] \hat{c}_{\rm{in}}^\dagger[\omega].
\end{aligned}
\end{equation}
The coefficients appearing in the above expression are
\begin{equation}
\label{eqapp:coeffs}
\begin{aligned}
\mathcal{G}_{x, \theta} [\omega] &=\sqrt{\kappa_c} G g_{mc} \left[ \chi_c[\omega] \Xi_m[\omega] e^{i (\varphi - \theta) } + \chi_c^*[-\omega] \Xi_m^*[-\omega] e^{-i (\varphi - \theta)} \right], \\
\mathbb{A}_{m;\theta} [\omega] &= i g_{mc} \sqrt{\frac{\kappa_c \kappa_m}{2}} \chi_c[\omega] \Xi_m[\omega] e^{-i \theta}, \\
\mathbb{A}_{c;\theta} [\omega] &=\left(1-\kappa_c \frac{\chi_c[\omega]}{\chi_m[\omega]}\Xi_m[\omega] \right)\frac{ e^{-i \theta}}{\sqrt{2}}.
\end{aligned}
\end{equation}

The correlation noise spectrum between two of such quadratures is given by
\begin{equation}
S_{\theta \theta^\prime}[\omega] = \int \frac{d \omega^\prime}{2 \pi} \langle \hat{x}_{\rm{out}, \theta}[\omega] \hat{x}_{\rm{out}, \theta^\prime}[\omega^\prime] \rangle,
\end{equation}
and then the self-correlation, obtained for $\theta=\theta^\prime$ reads explicitly
\begin{equation}
\begin{aligned}
S_{\theta \theta}[\omega] &= \vert \mathcal{G}_{x, \theta} [\omega] \vert^2 S^{(0)}[\omega] + S^{(0)}_{\rm{out}}[\omega] \\
&=\vert \mathcal{G}_{x, \theta} [\omega] \vert^2 S^{(0)}[\omega] + \vert \mathbb{A}_{m, \theta}[\omega] \vert^2(n_m+1) + \vert \mathbb{A}_{m, \theta}[-\omega] \vert^2 n_m + \vert \mathbb{A}_{c, \theta}[\omega] \vert^2(n_c+1) + \vert \mathbb{A}_{c, \theta}[-\omega] \vert^2 n_c,
\end{aligned}
\end{equation}
where $S_{\rm{out}}^{(0)}[\omega]$ is the output spectrum in the absence of the magnomechanical coupling. In order to obtain the noise added to the measurement of $S^{(0)}[\omega]$, we write
\begin{equation}
\begin{aligned}
S_{\theta \theta}[\omega] &= \vert \mathcal{G}_{x, \theta}[\omega]  \vert^2 \left[ S^{(0)}[\omega] + S_{\rm{imp}}[\omega]\right],
\end{aligned}
\end{equation}
where
\begin{equation}
\label{eqapp:impnoise}
S_{\rm{imp}}[\omega] = \frac{1}{\vert \mathcal{G}_{x, \theta} [\omega]  \vert^2} \left[ \vert \mathbb{A}_{m, \theta}[\omega] \vert^2(n_m+1) + \vert \mathbb{A}_{m, \theta}[-\omega] \vert^2 n_m +  \vert \mathbb{A}_{c, \theta}[\omega] \vert^2(n_c+1) + \vert \mathbb{A}_{c, \theta}[-\omega] \vert^2 n_c \right].
\end{equation}
We notice now that, at $\omega = 0$
\begin{equation}
\begin{aligned}
S_{\theta \theta}(0) &= \vert \mathcal{G}_{x, \theta}[0]  \vert^2 \left[ S^{(0)}[0] + S_{\rm{imp}}[0]\right] \\
&= \vert \mathbb{A}_{x, \theta}[0]  \vert^2 \frac{2}{\gamma_b} \left[ 2 n_b +1 + \frac{\gamma_b}{2} S^{(0)}_{\rm{imp}}[0]\right].
\end{aligned}
\end{equation}
We then define the added quanta to the measurement due to the imprecision noise as
\begin{equation}
n_{{\rm{imp}}, \theta} = \frac{\gamma_b}{4} S_{{\rm{imp}}, \theta}[0].
\end{equation}
Since for frequencies $\omega$ close to zero we have $S_{\rm{imp}}[\omega] \approx S_{\rm{imp}}[0]$, we can write
\begin{equation}
S_{\theta \theta}[\omega] = \vert \mathcal{G}_{x, \theta}[\omega]   \vert^2 \left[S^{(0)}[\omega] + \frac{4 n_{{\rm{imp}}, \theta}}{ \gamma_b} \right].
\end{equation}
We note that for schemes that break the BAE conditions, the phonon quadrature will include a backaction noise term, which then correlates with the imprecision noise.

The imprecision noise for a given angle $\theta$ is below the standard quantum limit provided that $n_{{\rm{imp}}, \theta} < 1/2$. For magnons at resonance with microwaves, the bath occupancy of such modes $n_{m,c}$ has to satisfy
\begin{equation}
n_{m,c}< \frac{\vert \mathcal{G}_{x, \theta}[0]  \vert^2}{\gamma_b (\vert \mathbb{A}_{m, \theta}[0] \vert^2) + \vert \mathbb{A}_{c, \theta}[0] \vert^2) } - \frac{1}{2}.
\end{equation}
The maximum occupancy of the magnon/microwave baths for beating the SQL depends on the gain factor $\vert \mathcal{G}_{x, \theta}[0]  \vert^2$, which scales linearly with the drive power. In general, strong drive powers allow to beat the SQL at higher bath occupancies, while smaller phonon decays also allow higher $n_{c,m}$. We should nevertheless notice that the typical temperature required for having a negligible occupancy of the magnon/microwave bath is of a few hundreds of mK, a condition attainable routinely in cavity magnonics experiments.

\section{Position noise spectrum: deviations from the BAE setup}

Starting with the full solution for $\hat{b}[\omega]$ given in Eq.~\eqref{eq:fullb}, we drop any term that has an argument containing $\sigma_b$ and solve the set of equations for $\hat{b}[\omega]$ and $\hat{b}^\dagger [\omega]$. This yields

\begin{equation}
\label{eq:solb}
\begin{aligned}
\chi_{b, {\rm{eff}} }^{-1}[\omega] \hat{b}[\omega] &= (i g_+^* \mathcal{B}[\omega]- i g_-^*)\hat{\xi}_{\rm{BA}}[\omega + \delta_b] + (i g_- \mathcal{B}[\omega]- i g_+ )\hat{\xi}_{\rm{BA}}^\dagger[\omega + \delta_b] \\
&\quad  + \sqrt{\gamma_b} \hat{b}_{\rm{in}} [\omega] + \sqrt{\gamma_b} \mathcal{B}[\omega] \hat{b}_{\rm{in}}^\dagger [\omega + 2 \delta_b],
\end{aligned}
\end{equation}
where we have defined the auxiliary function
\begin{equation}
\mathcal{B}[\omega] = \frac{f_2 [\omega]}{\chi_b^{-1} [\omega + 2 \delta_b] + i \Sigma_b^* [-\omega - 2 \delta_b]}.
\end{equation}
The effective phonon susceptibility $\chi_{b, {\rm{eff}} }^{-1}[\omega]$ is given by
\begin{equation}
\chi_{b, {\rm{eff}} }^{-1}[\omega] = \chi_b^{-1}[\omega] - i \Sigma_b[\omega] - f_2^*[-\omega - 2 \delta_b] \mathcal{B}[\omega].
\end{equation}
The self-energy term $\Sigma_b[\omega]$ is given in Eq.~\eqref{eq:phononselfenergyfull}, while the function $f_2[\omega]$ was defined in Eq.~\eqref{eq:f1f2}.

Using then the expectation values for the correlations of $\hat{\xi}_{\rm{BA}}[\omega + \delta_b]$, we can obtain the mechanical noise spectrum, whose integrals gives the results shown in the main text.

With Eq.~\eqref{eq:solb}, we also obtain the solution for the microwave mode operator $\hat{c}[\omega]$, which we use to construct the microwave output $\hat{c}_{\rm{out}}[\omega]$ via the standard input-output relation. The calculation of the output spectrum then follows the same procedure we used before, with the difference that the terms $\hat{\xi}_{\rm{BA}}$, appearing in Eq.~\eqref{eq:solb}, do not cancel, and thus the ouput spectrum has both backaction noise and noise correlation contributions. As in the case of a single tone drive, the noise correlation is antisymmetric in frequency, and thus the symmetrized noise spectrum does not have such a contribution.

\bibliography{apssamp}

\end{document}